# Challenges for Resistive Gaseous Detectors towards RPC2014[1]

**Diego Gonzalez-Diaz**[a,b,c,*], **Archana Sharma**[d]

[a] *Laboratorio de Física Nuclear y Altas Energías, University of Zaragoza c/ Pedro Cerbuna 12, 50009, Zaragoza, Spain*
[b] *GSI Helmholtz Center for Heavy Ion Research, Planckstrasse 1, 64291, Darmstadt, Germany*
[c] *Department of Engineering Physics, Tsinghua University, Liuqing building, 10084, Beijing, Germany*
[d] *CERN, Geneva 23, CH-121, Geneva, Switzerland*

*E-mail*: diegogon@unizar.es

**ABSTRACT:** Resistive gaseous detectors can be broadly defined as a sub-type of gaseous detectors that are operated in conditions where virtually no field lines exist that connect any two metallic electrodes sitting at different potential. For most practical purposes, this condition can be operationally realized as 'no gas gap being delimited by two metallic electrodes' [1]. Since early 70's, Resistive Plate Chambers (RPCs) are the most successful implementation of this idea, that leads to fully spark-protected gaseous detectors, with solid state –like reliability at working fields beyond 100kV/cm, yet enjoying the general characteristics of gaseous detectors in terms of flexibility, optimization and customization. We present a summary of the status of the field of resistive gaseous detectors as discussed in a dedicated closing session that took place during the XI Workshop for Resistive Plate Chambers and Related Detectors held in Frascati, and especially we review the perspectives and ambitions of the community towards the XII Workshop to be held in Beijing in year 2014.

[1] Text based on discussions between attendants to the XI workshop on Resistive Plate Chambers and Related detectors and a panel formed by S. Bianco, P. Dupieux, P. Fonte, G. Iaselli, A. Musso, S. K. Park, R. Santonico, V. Peskov.

## Contents

## 1. Introduction

More than twenty five years have passed after the conception of the Large Hadron Collider, back in late 80's [2]. At the moment of writing, at CERN, Geneva, the LHC is dramatically pushing our frontier of knowledge to energy regimes never explored before. This is bringing not only tantalizing hints in the long-awaited Higgs sector, the dense and hot partonic phase expectedly formed in very energetic heavy ion collisions is also being studied with higher-than-ever precision, as well as CP-violation [3].[2] There is an ongoing (so far unsuccessful) search for evidence of super-symmetric particles that might explain the dark matter needed by our current cosmological model, alongside a number of Standard Model extensions and other exotica.

Resistive gaseous detectors under the standard form of Resistive Plate Chambers (RPCs) [4,5,6] play a fundamental role in this global scientific breakthrough, due to their high-end timing properties, reliability, ability to easily scale to large areas and magnetic field immunity.

[2] At the time of submission of this work, both CMS and ATLAS collaborations have confirmed the existence of a Higgs-like particle with a mass around $M_H$=125MeV/$c^2$.

Amongst other advantages, which are partly addressed in this document, they offer a rational option for the fast triggering and tracking of muons over large areas (for a review see [7]).

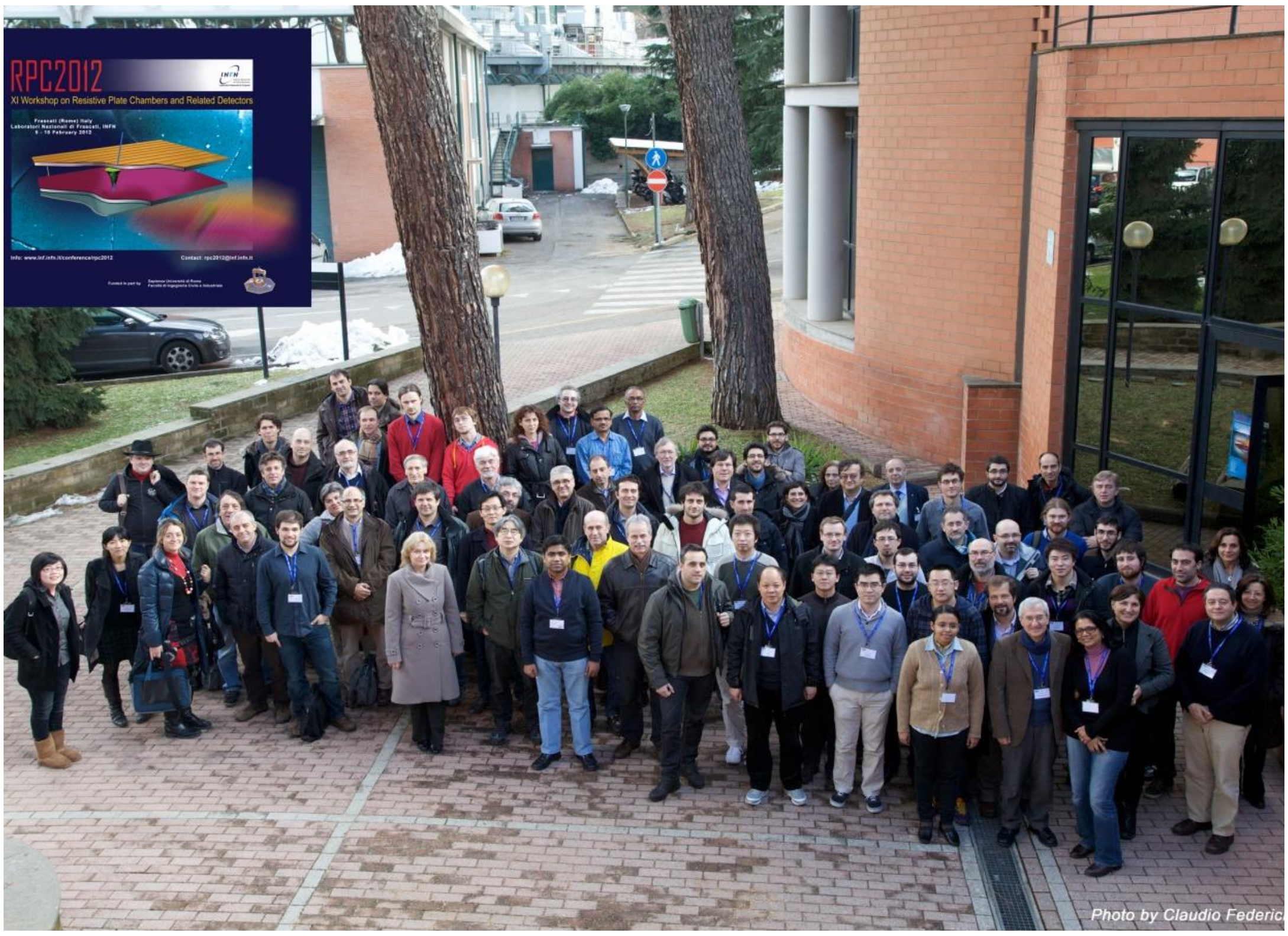

**Figure 1:** Official photo of the attendants to the XI Workshop on Resistive Plate Chambers and Related Detectors.

The technological branching that took place in the second half of the 90's [8, 9] has sped up the dissemination of the RPC technology into two main directions:

i) *classic single-gap RPCs*, like the ones constituting the bunch-crossing taggers, namely, the trigger systems of CMS and ATLAS [10,11] (an optimal and well-developed choice for the coverage of large areas when aiming at scintillator-like reliability, magnetic field immunity and ~1-2 ns time-of-flight tagging accuracy, at high efficiency), and

ii) *multi-gap timing RPCs* for particle identification purposes, like the ALICE TOF-barrel [12] (offering also high reliability and magnetic field immunity, but intrinsically more versatile and capable of providing sub-100ps time-of-flight tagging accuracy, also at high efficiency).[3]

[3] Presently, RPCs are also naturally classified from a structural point of view as ‘wide gap’ (~1mm) or ‘narrow gap’ (<0.3mm), or functionally as ‘trigger’ or ‘timing’, depending on the context, [1]. We follow in this document a technological classification, so as to acknowledge the technical merit of the RPC [4, 5, 6], multi-gap [8] and timing [9] developments. There is little ambiguity at the moment, though, whatever adjective is used.

From the early days of the *Pestov counter* back in 1971 [4, 5], there has been indeed a large diversification and growth of the RPC field, especially once the technology developed by the soviet school was adapted to commercial materials and made workable at atmospheric pressures yet with high efficiency to minimum ionizing particles, a crucial step accomplished by Santonico and Cardarelli in 1981 [6]. The term ‘Resistive Plate’ was hence coined, giving birth to the ‘Resistive Plate Chamber’ (RPC) technology. This breakthrough was to result in a slow but steady democratization of the usage of gaseous parallel plate configurations worldwide. Some consequences of this evolution we can see today, exemplified by the broad range of resistive-plate choices, the universal operation at atmospheric pressure, and the reduction of the initially very extreme operating conditions down to a much more comfortable (yet extreme) saturated-avalanche mode, with a large part of the amplification load moved into the front end electronics. The advent of the multi-gap configuration in 1996 [8] finally allowed to technologically ‘decouple’ the requirements for efficiency and time resolution, thus allowing to obtain both excellent timing and efficiency even at atmospheric pressure. Technical and practical problems apart, the time resolution is a matter of the gap characteristics, while a multiple gap can always ensure enough creation of primary ionization clusters, through a procedure that is nowadays known to be easily scalable (‘multi-gap’). Not having found yet the point where these two ideas conflict, the ultimate timing limits of the RPC technology remain widely open (see [13], for instance).

During the arrangements for the organization of the XI Workshop on Resistive Plate Chambers and Related Detectors (Fig. 1) it was felt that, in view of the recent developments and growth of the RPC technology, a comprehensive compilation of the status and the main future challenges in the field was needed and, with that purpose, the present contribution was born. It is organized along substantial topics discussed during the workshop closing session, which took place around a Discussion Panel constituted by experts in the field and that was moderated by the authors, as follows:

1. Introduction.
2. Status of classic and multi-gap timing RPCs in running experiments.
3. Gas-related ageing, gas chemistry, gas availability and usability.
4. Rate capability, material research and modeling, long-term stability of the electrical properties of the resistive plates.
5. Electronic developments. Discrete electronics and general-purpose ASIC-based amplification/discrimination boards. High accuracy Time to Digital converters.
6. Status and future of simulations.
7. Future of classic and multi-gap timing RPCs.
8. New applications.
9. New trends in resistive gaseous detectors.
10. Conclusions.

## 2. Status of classic and multi-gap timing RPCs in running experiments

### 2.1 Classic RPCs

Being originally conceived about two decades ago, CMS and ATLAS sub-systems based on classic RPCs are, however, not first of a kind. Fully fledged large-scale physics experiments, importantly the B-factories BaBar [14] and Belle [15], used them as muon triggers way before LHC started producing its first scientific output, albeit chambers were operated in streamer mode (see later). Both CMS and ATLAS RPCs, each extending over a breathtaking area of 4000 $m^2$, have removed all doubt concerning the technologically subtle process of linseed oil coating, which is needed for low dark current and stability of operation of classic RPCs in avalanche mode. Fig. 2 shows a 3D-view of the intrinsic efficiency of one of the CMS barrel wheels as obtained for proton collisions at a center of mass energy of $\sqrt{s}$=7 TeV. Values well above 95% (average 96.5%) are comfortably and uniformly reached over the whole system [16]. Efforts currently focus on the purification of the gas system, sophisticated humidity monitoring (needed to stabilize the conductivity of the Bakelite/HPL used as resistive plate), background characterization and on-line *E/p* compensation. Taking into account the soft (nearly linear) dependence of the gain with the reduced field that is characteristic of operation in saturated-avalanche regime, the latter correction can be taken as a good indicator of the level of sophistication that the technology is reaching (efficiency after correction is stable with operating time within a remarkable 0.5%, i.e., 1/200 particles [16]).

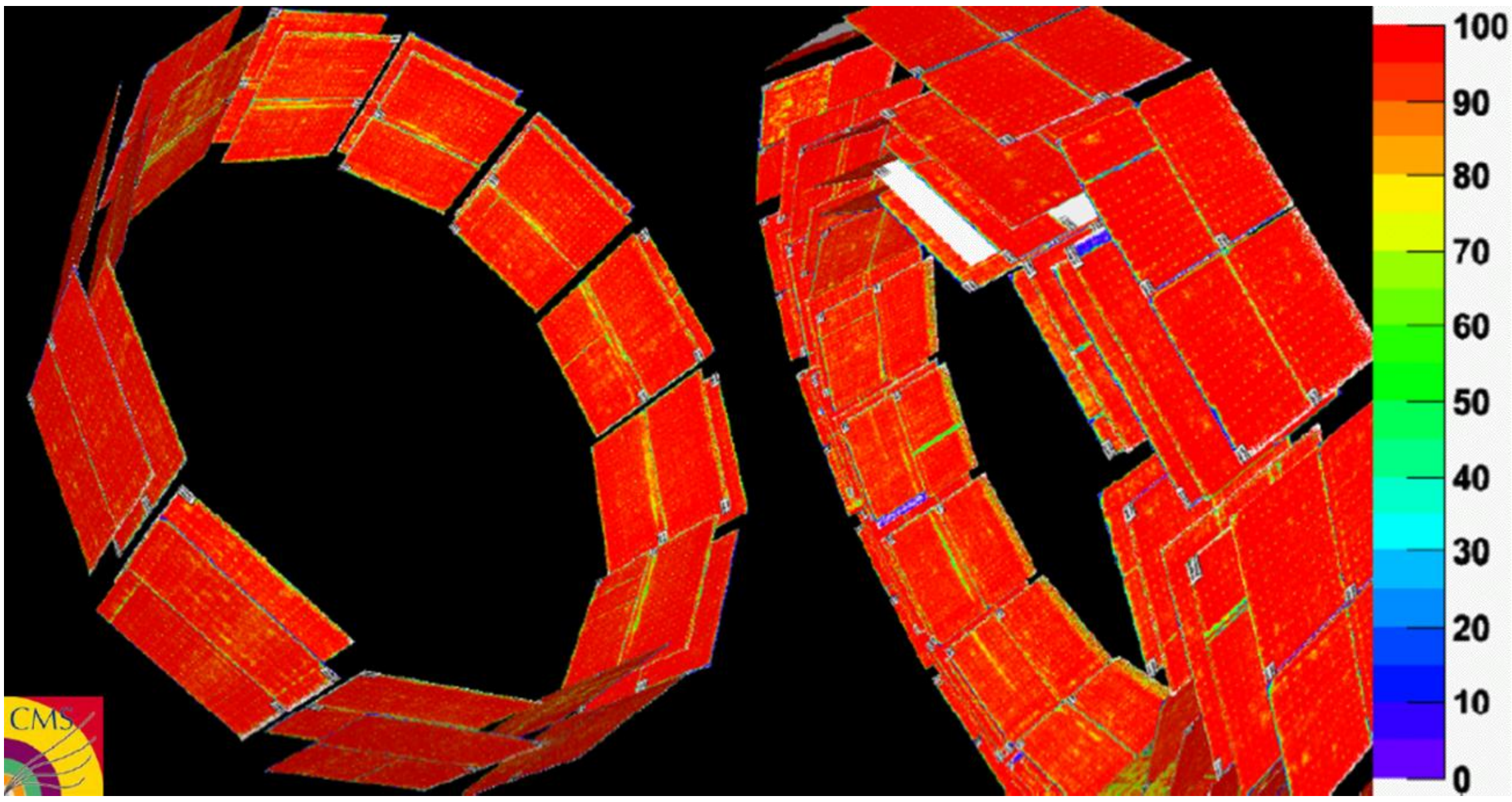

**Figure 2.** The two innermost radial layers (left) and five radial layers, except chamber RB4, (right) of the classic RPCs constituting one of the CMS barrel wheels. The 3D color-code representation depicts the chambers efficiency after online compensation of environmental variations of the reduced field E/p, with axis given to the right, [16].

On another hand, there has been a recent interest in exploiting the timing of classic RPCs down to the one intrinsically expected ($\sigma_T$ ~1.5-1.75 ns in ATLAS), in view of the need for extending the RPC capabilities beyond its prime role as bunch-crossing tagger.[4] After a dedicated calibration, mainly involving the correction of system delays, an overall time resolution down to $\sigma_T$ ~1.99 ns has been obtained (Figs. 3). This achievement has allowed providing some (limited) particle identification capabilities, in particular resulting in the suppression of loopers (particles trapped in the magnetic field), unphysical beam collisions, and natural background radiation [17]. Despite the progress, there is the belief that the full potential of both CMS and ATLAS RPCs has not being fully exploited yet [18].

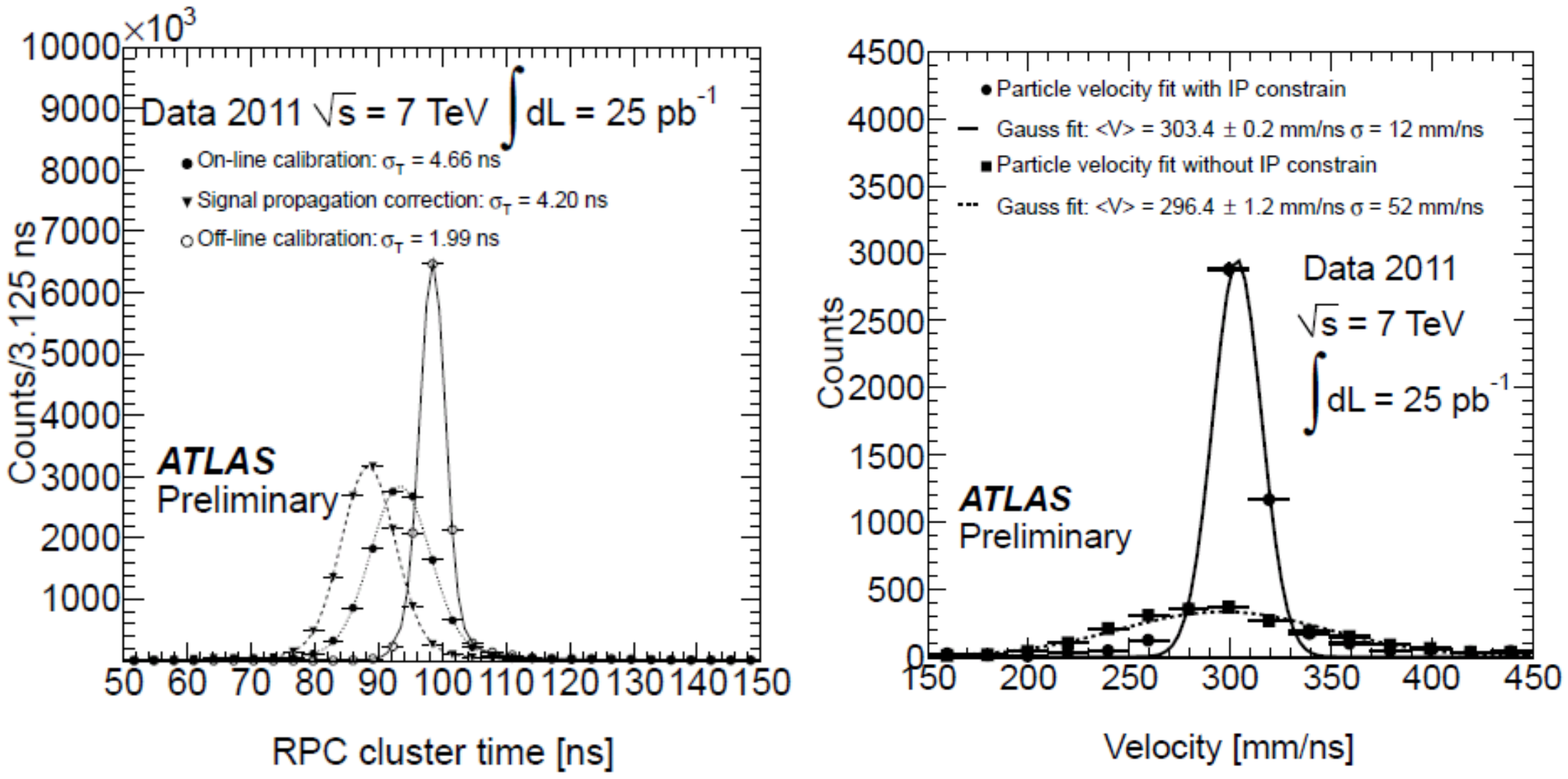

**Figure 3.** Left: time resolution from the ATLAS spectrometer after off-line calibration. The observed resolution is interpreted by authors as having a contribution of 1.5 ns from the counters, 0.9 ns from the 3.125 ns sampling clock, and approximately 1ns from the remaining electronics chain. Right: reconstructed velocity distribution with an interaction point (IP) constraint (circles) and without it (squares). Figures taken from [17].

It turns out that classic RPCs provide also a remarkably simple technological solution for the coverage of large areas when merely aiming at accurate particle counting. Indeed, LHC-inspired RPCs (inheriting many of their architectural properties) running in the absence of streamer quenching (‘streamer mode’), hence with a much simplified electronic stage and limited time resolution, are optimally suited for the task. Such types of systems are already making contact with fundamental issues in both astrophysics (ARGO- cosmic rays, [19]) and particle physics (OPERA- neutrino oscillations, [20, 21]). As compared to ARGO and OPERA, the 140 $m^2$ ALICE muon spectrometer represents a slightly different use-case of the streamer mode operation. It aims at tagging the J/ψ and Υ production in heavy ion reactions via proper $p_T$ selection in their di-muon decay channel. A preliminary analysis of the J/ψ-suppression factor, determined in this way, has been presented at the workshop [22]. Other examples of muon trigger systems operated in streamer mode, some already mentioned at the introduction of this sub-section, are BaBar [14], Belle [15] and, more recently, BesIII [23].

[4] There are different usages of the concept ‘intrinsic resolution’ in literature. We refer, in this work, to a value for the time resolution that can be independently obtained under reference conditions where the relevant calibrations are either much simpler or straightforward. It thus represents a reasonable practical limit in light of existing evidence.

Although not addressed at the workshop, it is remarkable the recent usage of classic RPCs in streamer mode as veto/tracker systems, as for instance in the reactor neutrino experiment DayaBay [24]. These developments are extremely competitive with either solid state alternatives (plastic scintillators) or other standard gas-based detectors not being spark-protected. One can therefore expect this trend to continue in the near future.

### 2.2 Multi-gap timing RPCs (MRPCs)

Pretty much like with classic RPCs, original designs in multi-gap fashion of presently running time-of-flight systems (ALICE [12], STAR [25], HADES [26], FOPI [27]) were conceived more than a decade ago. Fully fledged systems became available and dismantled meanwhile in some singular cases, as HARP [28], showing very respectable performances, yet far from the ultimate potential of the technology. Present systems, either with pad [12, 25], multi-strip [27] or single-strip [26] readout, are already capable of comfortably reaching the 100 ps landmark and better (see Fig. 4 for a FOPI particle identification plot). Current efforts in HADES [29] and ALICE [30] are focused on the calibrations needed for obtaining, at system level, a time of flight resolution down to the intrinsic one (~70 ps and ~50 ps for minimum ionizing particles, respectively). A time-of-flight spectrum for Pb-Pb collisions at $\sqrt{s_{NN}}$=2.76 TeV as obtained by ALICE is given in Fig. 5. For the corresponding particle identification plot the reader is referred to [30].

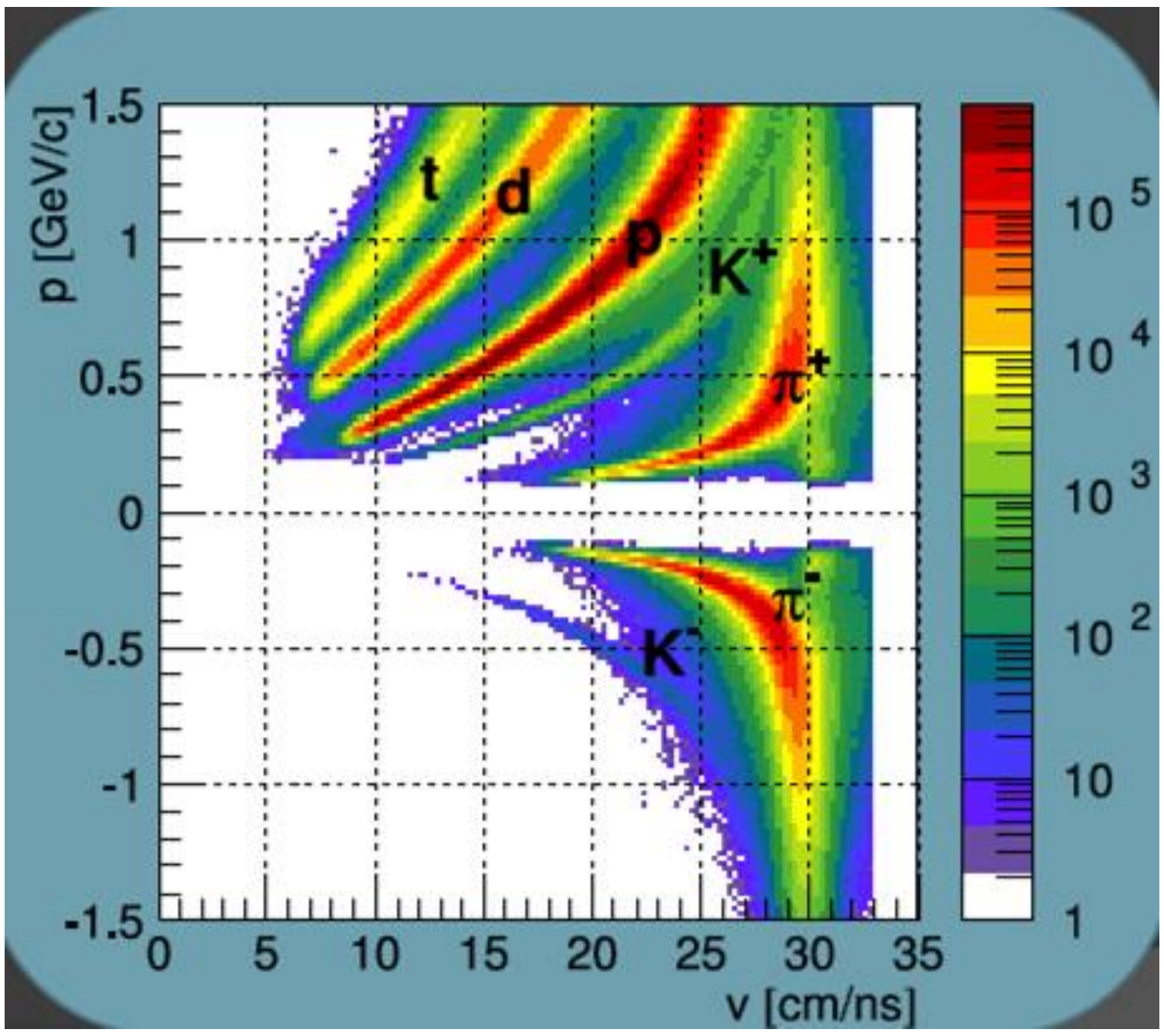

**Figure 4.** Particle identification plot from the FOPI experiment as obtained at SIS-18 energies, using multi-gap timing RPCs. $K^-$ dynamics sub or close to threshold is a key and difficult observable in heavy ion reactions in the energy regime $E_{lab}$ = 1-3 GeV/A; here they are visible to the naked eye. The plot is compatible with an overall system resolution of $\sigma_T$ = 90 ps, including the start detector.

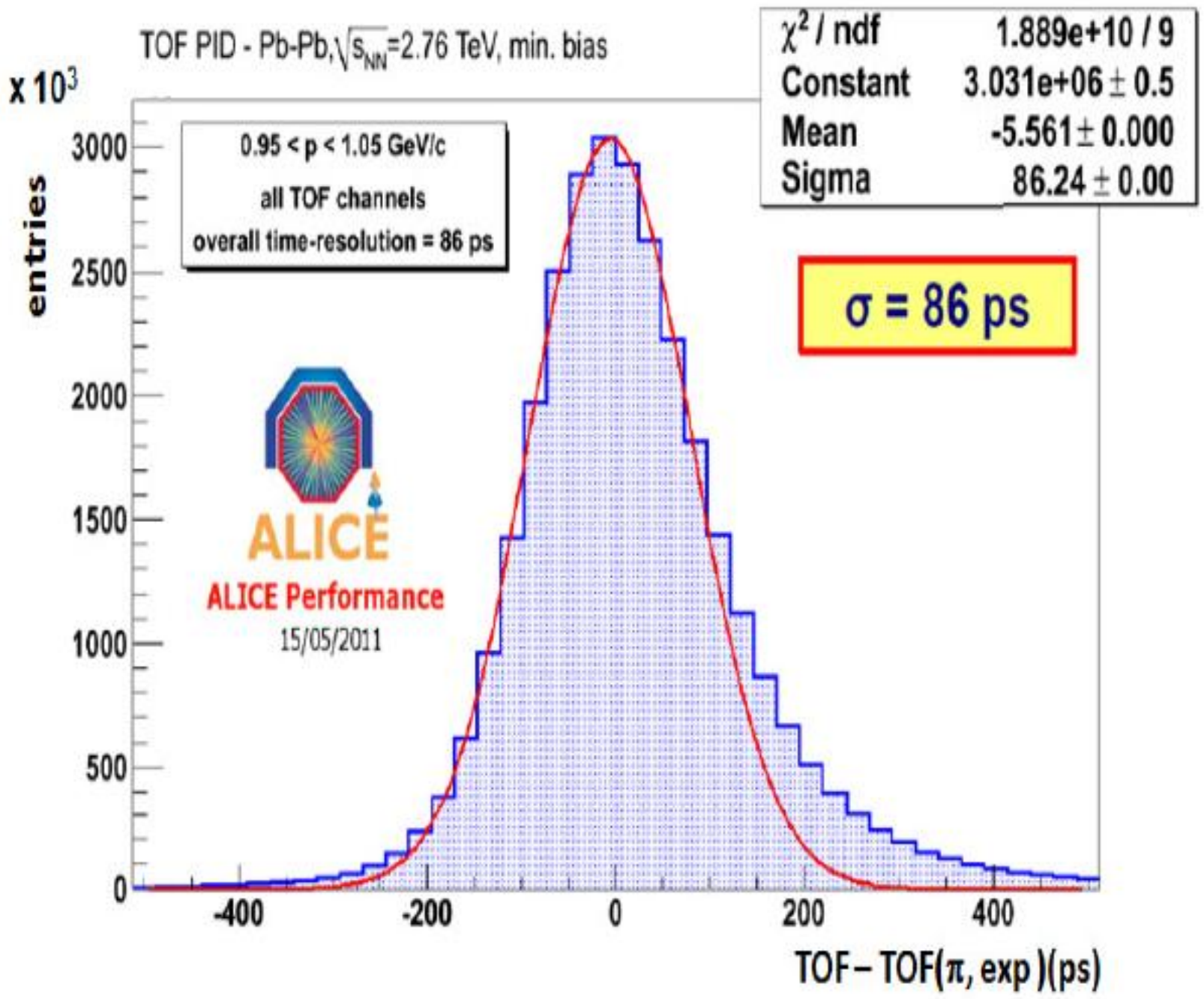

**Figure 5.** Time of flight distribution measured by the ALICE collaboration for selected pions, stemming from Pb-Pb collisions at $\sqrt{s_{NN}}$=2.76 TeV (from [30]). The collaboration aims at approaching the 50ps landmark with more advanced analysis.

Remarkably, a novel MRPC project has recently emerged for large-area cosmic ray studies at ground level, within the EEE collaboration. Despite the project's primordial interest is educational, some interesting scientific output has been already produced (see [31] and references therein). The collaboration has managed to obtain 100 ps-level resolutions in some counters; however the impact of this performance on the event reconstruction and data analysis has not been explored yet.

In all, there is a general consensus that the expectations (so far) raised by these new devices, namely, reliable operation over large areas at sub-100 ps timing and high efficiency for minimum ionizing particles, have been fulfilled.

## 3. Gas

In view of their large global warming potential (GWP), the usage of two of the main RPC gases, $C_2H_2F_4$ and $SF_6$, is progressively being banned by authorities ([32], for instance). The future importance of these two gases for the greenhouse effect cannot be neglected since, over a 100 years time-span, predictions give a GWP of 1300 for $C_2H_2F_4$ while $SF_6$ will exceed 20000 [33]; that, said, means being 20000 more dangerous than $CO_2$. There are two lessons to be learned from the past, based on the fact that these gases are massively used in industrial processes: 1) regulations might be flexible regarding usage for scientific purposes, 2) industry will tend to replace this gas by a functionally equivalent one, expectedly not worse (similar to the $CF_3Br$ case, [34]). The latter point is even more appealing in the case of $SF_6$: being one of the best inert electron scavengers known in nature, it therefore exhibits a property that is arguably of comparable importance for RPCs and for industrial HV insulation. Re-circulating systems are definitely an option worth being explored [33, 35] although for small R&D setups they might be not too economical. One could rely on a combination of the two aforementioned hypothetical circumstances, i.e., flexibility of regulations and/or industrial replacement for small R&D setups together with a robust closed loop operation for large systems. Although these options will definitely help avoiding 'sudden extinction' [36], a medium-term strategy seems to be unavoidable.

The most recent exploration of new gas mixtures was carried out by Abbrescia [37], who studied the possibility of adding Helium to the mixture as a 'space holder' gas (in the author's words), i.e., a gas that behaves roughly like a homogeneous vacuum, thus effectively reducing the gas density and consequently the necessary operating field (a fact that is generally considered to be advantageous, in a purely technical basis). The principle works, at least qualitatively. However the presence of the standard components of the mixture was respected and just their proportions varied a fact that does not really help at solving the problem (although it clearly alleviates it, at least partly). In any case, this effort can be very worthy considered as a first of a new set of experimental studies designed to eventually replace $C_2H_2F_4$ and $SF_6$ from the standard RPC mixture. Rather complementary, a recent systematic study of the role of both i-$C_4H_{10}$ and $SF_6$ in timing RPCs can be found in [38]. It should be stressed that $SF_6$–free operation is definitively possible for timing applications, and both the STAR barrel as well as the future STAR Muon Trigger Detector (MTD) operate in this mode in order to ensure safe operation of the central TPC. Some compromise must be accepted then in terms of a reduced plateau, higher presence of streamers and the subsequent degradation of time resolution. In practice, operation at 90-95% efficiency with resolutions in the range 90-110ps under $SF_6$–free gas mixtures has been convincingly demonstrated up to 1m-long strip counters [39].

Thus, on the one hand, the replacement of both $C_2H_2F_4$ and $SF_6$ seems to be unavoidable in the medium term. On the other hand, if the gas is to be replaced, ensuring low ageing for future experiments expecting to deal, over their operating life, with transported charges up to 1-3C/cm$^2$ [40, 41] will require of a systematic and dedicated approach to the problem. Only then it will be possible to guarantee the previously observed low ageing levels present in both classic and multi-gap timing RPCs operated in avalanche mode. At any rate, exploring new gas mixtures while pursuing a microscopic understanding and modeling of the gas-related ageing phenomena (along the promising lines in [42], for instance) will certainly require of a multi-disciplinary approach and several years of development, in a realistic scenario.

## 4. Material research

In the immediate future, RPCs able to withstand particle fluxes up to $\Phi$=20kHz/cm$^2$ are necessary both in view of the sLHC upgrade at CERN, Geneva, and of the future SIS-100/300 at FAIR, Darmstadt. By increasing the electronics sensitivity in a factor x5-10 (see next section), operation of classic RPCs has gone down to average charges per gap of around $Q$=2 pC in order to fulfill these rate capability requirements. This is a standard working number for multi-gap timing RPCs ([43], for instance), thus it is not surprising that both CBM MRPCs (at FAIR) and ATLAS classic RPCs (at sLHC) are planned at the moment based on materials of similar bulk resistivity ($\rho_{20}$~$10^{10}$ Ωcm). Such a remarkable technological convergence arises from a common necessity of minimizing the dynamic drop of the operating field at high particle fluxes, stemming from the ohmic voltage drop at the resistive plates ~ $\rho Q\Phi$, down to tolerable levels. Despite the similarities, the materials employed have a priori a fundamental difference: Chinese doped glass (one of the options for the CBM experiment) is polished to ensure the high surface quality required for timing while ATLAS relies on the well established technology of linseed oil –coated Bakelite. General considerations on gas ageing apart (see previous section), ensuring the stability of the electrical properties of the plates is of utmost importance, up to transported charges in the order of 1-3 C/cm$^2$ (i.e., 5 years at the maximum expected flux and 50% duty cycle) and this has to be consistently demonstrated under realistic conditions. At present, promising results in DC(static)-conditions are available for Chinese glass up to 1 C/cm$^2$ (i.e., for a raw piece of material on the bench) and up to 0.05C/cm$^2$ under X-ray irradiation (i.e., once the chamber is assembled), showing no sizable effects [41]. On the other hand, since the ATLAS-upgrade RPCs will roughly reduce the operating charge per gap by the same amount as the flux will be increased, it can be reasonably expected a low ageing figure, this is to say, not worse than the one of the already commissioned system over a comparable operating time.

The challenge of finding new resistive materials was neatly addressed during previous workshop [44]. Broadly speaking, it can be understood as a lack of industrial interest in developing 'bad insulators' (Fig. 6), a definition corresponding precisely to the range where both adequate spark-quenching and adequate rate capability can be simultaneously obtained in parallel plate geometries. Besides Bakelite and float glass, materials in the range $\rho \sim 10^7$-$10^{13}$ Ωcm have to be custom-developed or are simply rare and/or expensive.

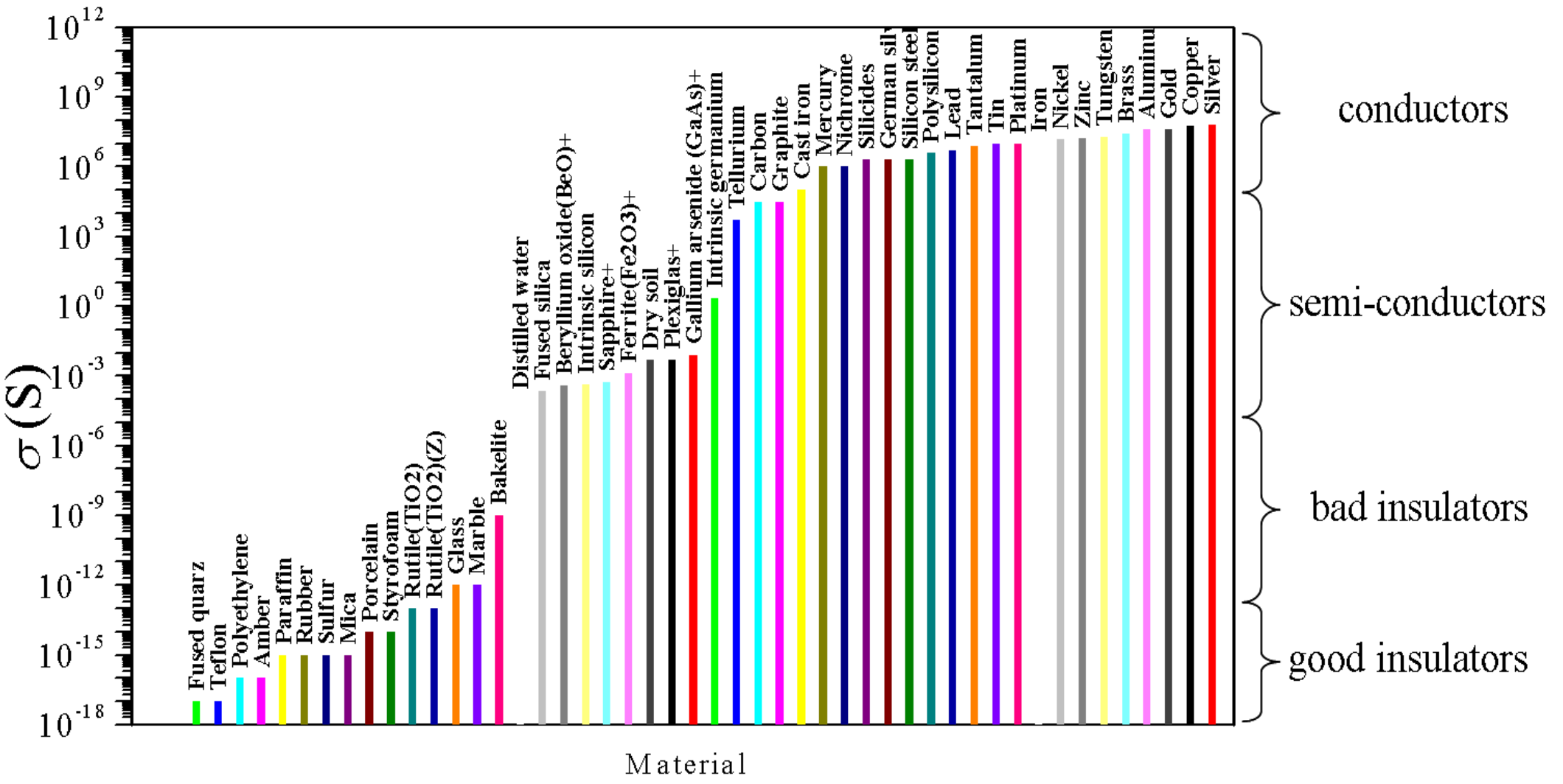

**Figure 6.** Modified figure from [44] showing a list of popular industrial materials spanning over 30 orders of magnitude of their characteristic electrical conductivity $\sigma$ (expressed in Siemens). The labeling of two regions as 'bad insulators' and 'good insulators' represent the authors' choice and does not correspond to any universally accepted terminology. The region labeled as 'bad insulators' is the region of interest for RPCs. Over more than 6 orders of magnitude only glass, marble and Bakelite can be found to have adequate conductivities for application to RPC architectures.

Fortunately, in the context of the CBM R&D for reaching rate capabilities up to 20kHz/cm$^2$, two new materials have emerged in this 'resistivity gap' and consistent results with prototypes have been shown for already few years: the aforementioned Chinese doped glass developed by Tsinghua University at Beijing [41, 45] with $\rho_{20}$~$2x10^{10}$ Ωcm (ohmic) and the $Si_3N_4$/SiC ceramics developed by the HZDR institute at Dresden [46], with $\rho_{20}$~$7x10^9$ Ωcm (varistor-type, tunable). The former has a smooth 'float glass'-like behavior with low dark current while the latter, allowing for a higher rate capability, has raised some concerns related to the small signal pick-up observed on large-area prototypes as well as large dark currents that are, as yet, unexplained. There are, however, strong indications that the rate capability of the $Si_3N_4$/SiC-based counters may comfortably reach the 100 kHz/$cm^2$ landmark under uniform irradiation, at least for small detectors.

Although, importantly, the modeling of the role of the electrical conductivity of the resistive material in the RPC behavior has been the subject of some interest in the past ([47-51]), a quantitative description of the fundamental process driving the conductivity was never attempted before, to the best of the authors' knowledge. Having such a model in hand will help enormously at understanding the limitations of certain approaches (in the most typical cases, resulting from instabilities appearing as a function of the operating time / transported charge density); it will prove also a very valuable tool for precisely defining the material requirements when, for instance, contacting partners from the field of material sciences or industry. The well-known failure at reproducing Pestov-like glass in Europe should be taken as a warning in this respect. Again, a multi-disciplinary approach has brought the most promising results and a model seems to be now within reach, at least for certain families of materials (Fig. 7). A consistent conductivity model can be expected to describe the behavior as a function of the electric field, temperature, time, and transported charge density under the same formalism [52]. Developments in this direction have an enormous potential in applied research ([53, 54], for instance) and should be prioritized.

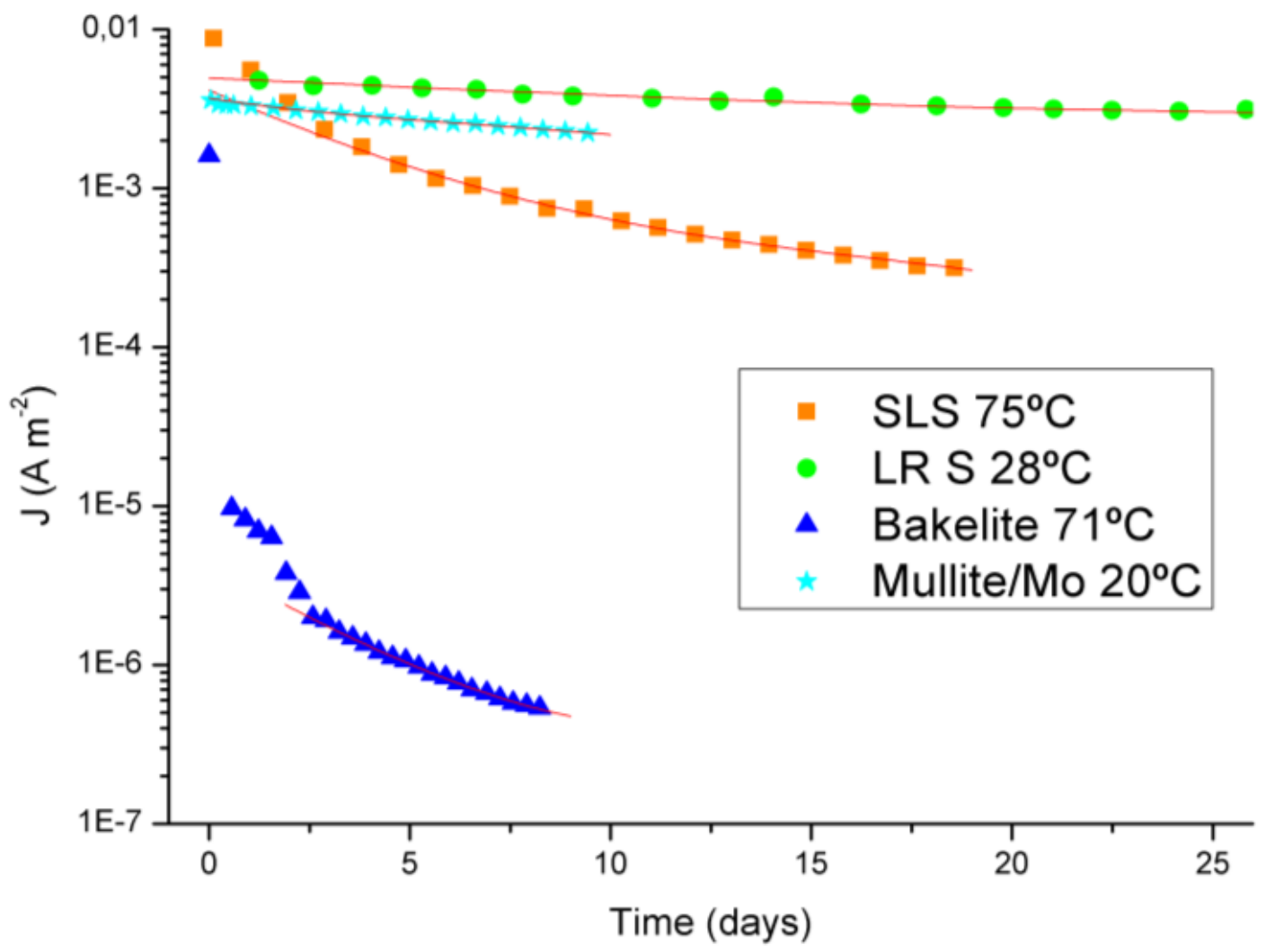

**Figure 7.** Behavior of the electrical current density flowing across different material samples for a constant voltage difference applied in a parallel plate configuration (~1/ρ, i.e., inverse to the sample resistivity) for different materials (soda-lime silica-glass -SLS, Chinese low-resistivity doped glass -LR, Bakelite and a novel Mullite/Molibdenum ceramic-metal composite). Lines show the behavior based on a conductivity model proposed by Hyde and Tomozawa (1986), (taken from [52]).

In the context of this and previous section, the elaboration of a list of RPC-compatible materials will be increasingly necessary, especially now that operating particle fluxes are going to see a 10 to 100-fold increase.

## 5. Electronics developments

The importance of the front end electronics (FEE) was very much stressed at the workshop. This fact obeys a 4-fold reason:

I) First of all, the coverage of large areas with high position resolution (on the order of 100-200μm) as is typical of applications in medical imaging or homeland security (see section 7) can hardly run with discrete components due to practical limitations as well as power consumption and space. So, the development of ASICs is an appealing direction to go in order to further explore these possibilities. Some fairly general-purpose broad-band amplifier/discriminator ASICs like NINO [55] and PADI [56] have been developed and are already under steady use, while a new ASIC, CAD, was presented at the workshop [57].

II) As mentioned in section 4, a reduction on the operating threshold together with the addition of 1-gap ('bi-gap') has allowed the ATLAS detectors to reach operating particle fluxes of 10 kHz/cm$^2$, at an average charge released per gap of around ~2 pC, close to the requirements of the ATLAS upgrade [40]. Thus, increasing the FEE sensitivity has allowed for a x5-10 reduction in the operating gain (Fig. 8). This fact results on a power dissipation for 10 kHz/cm$^2$ comparable to the one the present ATLAS RPCs have at the efficiency plateau for a meager 1-2 kHz/cm$^2$. As said, the satisfactory ageing figure of the already commissioned system is hence expected not to be worse with the new architecture. This idea of 'moving' gain from the detector to the FEE suggests that treating the FEE and RPC as separate entities is not the correct approach for present and future developments [58].

III) It is well known that current amplifiers show significant performance differences as compared to classic charge-sensitive amplifiers [59, 60]. With approximate character, the difference can be traced back to a bandwidth reduction of the latter and higher signal to noise ratio. This fundamental difference is, however, not properly addressed by existing simulation works and techniques when aiming at a quantitative description of the detector response properties (see next section). Until the electronics and detector are not treated consistently under the same simulation framework, it can be expected that achieving a very necessary global system description will never be fully achieved.

IV) The idea of developing some general-purpose electronics for the RPC community was addressed at the workshop. It was acknowledged that it would be, in general grounds, a very good idea. The key issues are the time jitter:

$$\sigma_T = \frac{\sigma_V}{dV/dt\,\big|_{V=V_{th}}}$$

and minimum threshold, say $V_{th}$ ~3-4$\sigma_V$. It is unfortunate that the slope at threshold *dV/dt* depends critically on both the signal shape and the amplifier response function, while the noise r.m.s., $\sigma_V$, and the overall stability are very much system-dependent. It will take some well-educated systematic study to successfully bridge the latter points and come to the specifications of such a general-purpose electronics. On the other hand, it is foreseeable that there will be always special applications that pose special requirements, for instance in terms of multi-hit response (i.e., cross-talk) or signal sensitivity. The complexity of this endeavor, therefore, can not be taken lightly.

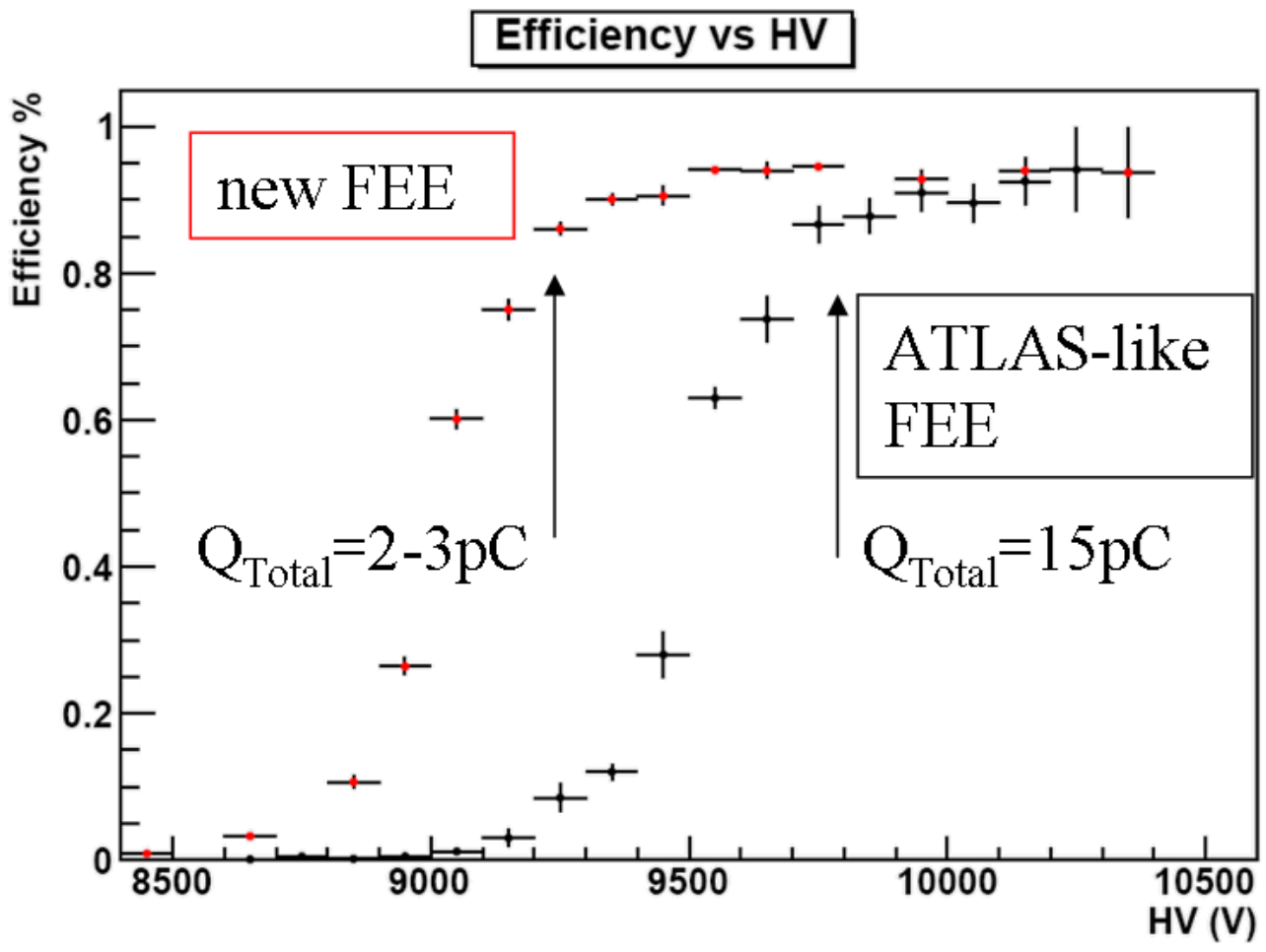

**Figure 8.** Comparison between the present ATLAS FEE and the newly developed one. The average charge released per amplification gap (here labeled with sub-index total, to distinguish from the electron-induced prompt charge) can be reduced by a factor 5-10, in virtue of the much increased electronics sensitivity.

Besides the topic of front end amplifying electronics, there has been considerable progress in the near past on digitization ASICs, most remarkably the HPTDC [61]. Now, a new FPGA-based board capable of reaching an unprecedented sub-10ps resolution has been presented at the workshop [62] (Fig. 9). It will be very exciting to see how this high-precision quest evolves and in particular if the RPC technology can keep pace.

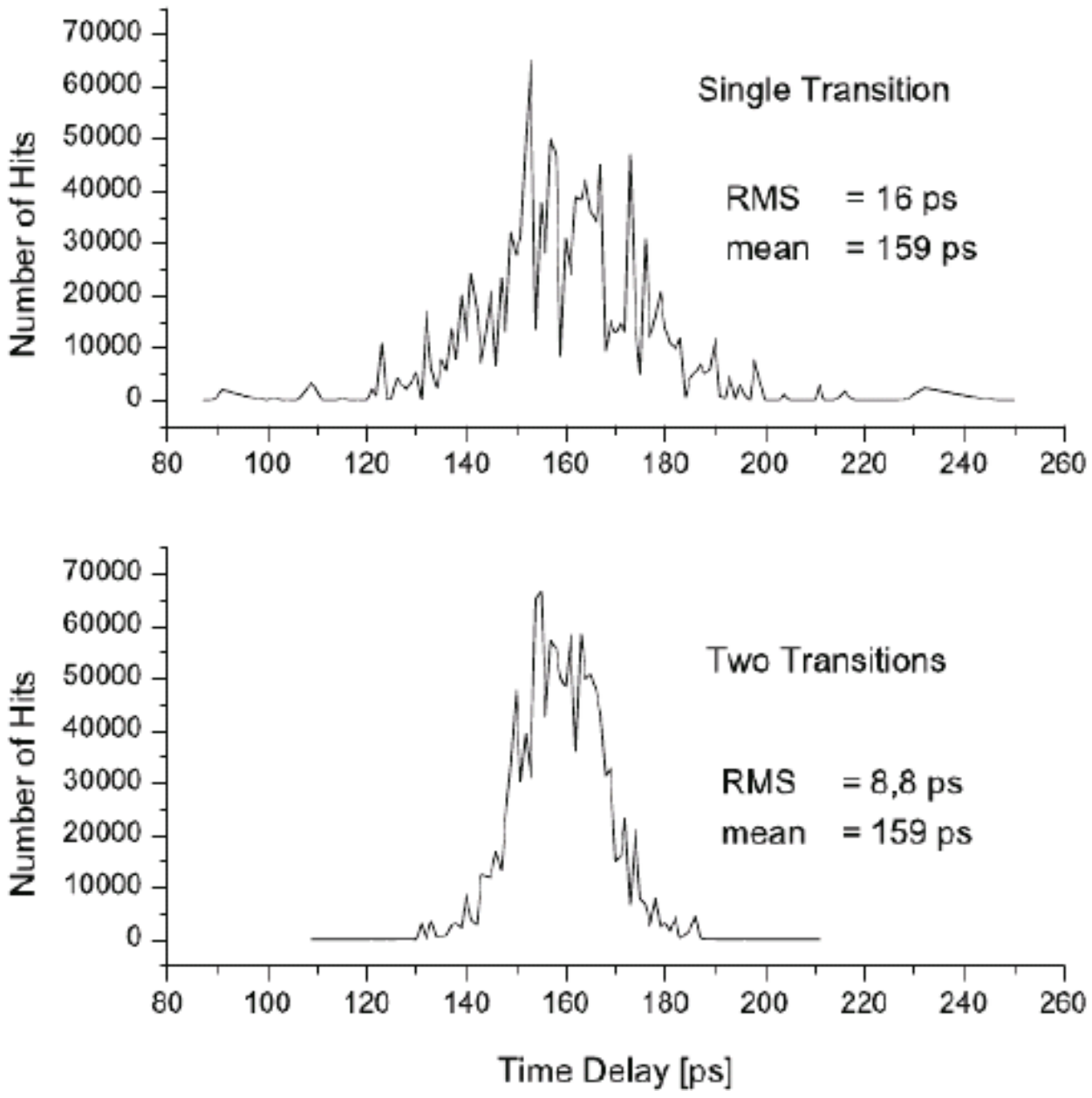

**Figure 9.** Sub-10ps time resolution obtained with an FPGA-based design, after processing 2 digital signals created by a fast pulse generator [62].

## 6. Status and future of simulations

### 6.1 The hydrodynamic model

A simulation mile-stone was achieved in 2004 with the description of the avalanche evolution in Resistive Plate Chambers under its own space-charge, largely utilizing first principles [43]. The simulations resorted to the minimal framework for realism, the 1.5D hydro-dynamic model [43, 63], that allows for a realistic calculation of the electric field during the avalanche evolution, yet within a reasonable computation time. In order to approximately account for the fluctuations in the multiplication process, the equations of evolution were solved in Monte Carlo (MC) fashion through a particle in cell (PIC) -like procedure. Thanks to improvements in the knowledge of the parameters of the gas (mainly in the initial ionization and multiplication coefficient), the simulations convincingly explained both the high experimental efficiencies and the charge spectra, that turn out to be completely dominated by the single avalanche self space-charge dynamics. Despite the success of the 1.5D model, a discrepancy up to a factor of 2 was observed in the description of the charge spectra (either prompt or total) while the parameters driving the electronics response function were tuned to describe the observed efficiency and time resolution data, albeit under not unreasonable assumptions. Importantly, much simpler descriptions based on 1D hydro-dynamic models achieve, overall, a higher level of accuracy; they require, however, of one extra parameter to be adjusted, namely, the number of carriers for which space-charge effectively sets in [64, 65]. On the opposite end, first principle 2D or 3D hydrodynamic MC simulations increase the computation time dramatically and are not wide-spread at the moment (for a comparison between 1.5D and 2D see [43]).

Thus, how to achieve a higher descriptive (and predictive) simulation power with potential to boost the parallel-plate technology further?. From the point of view of the avalanche dynamics alone, a natural step is to follow up with the hydro-dynamic approach in the 1.5D or even 2D scenario, where the computational overhead might still be acceptable for detector studies. But, first, a thorough revision of the gas parameters that characterize the electron swarm is pertinent. Exemplarily, early estimates of the multiplication (Townsend) coefficient from the Magboltz code (circa 2004), that is generally acknowledged to represent the state of the art for gaseous detector simulations, differ by as much as 20% from the recently measured values on the main RPC gas, $C_2H_2F_4$ [66] (see [65] for a comparison). According to [67], contemporary versions of Magboltz (beyond 8.X.X) have been tuned in order to describe the systematics measured in [66], although there is work still ongoing on this direction. It will be very interesting to see how the conclusions of previous simulation works are modified with much more precise parameters of the swarm, at least for a simple $C_2H_2F_4$—only scenario (gas mixtures add another level of complexity to Magboltz calculations, due to the presence of new hetero-nuclear atomic and molecular processes that are not present in the pure gases).

Importantly, the ultimate technological limits of the parallel plate technology are connected to the formation of streamers at the highest fields. Although in RPCs they do not manage to trigger breakdown, a deterioration of the time resolution is routinely observed when they are present in a sizable fraction, besides the consequent deterioration in rate capability and eventual appearance of undesired after-pulses. Part of the observed deterioration is certainly coming from the presence of precursors. Given the complexity of the streamer phenomena, only 1D hydro-dynamic continuous models have been used so far for describing the behavior of streamers in gaseous detectors [68]. A very qualitative application of this model to RPCs is available, through [69], where indeed the aforementioned experimental observation of streamer-precursors (for instance, [70]) has been reproduced. Further progress on this crucial aspect is connected both to the improvement in the knowledge of critical parameters of the gas (in particular, those connected to photo-production and photo-ionization) as well as the physical framework.

A major shortcoming when resorting to a hydrodynamic model (either in continuous or Monte Carlo fashion) is the fact that, due to the macroscopic character of the approach, it is necessary to give up any first-principle description of the avalanche statistics (of paramount importance at threshold level). This information is thus largely unknown at the moment both on the experimental and theoretical side, for the case of RPC gas mixtures. Despite some theoretical guidance on this aspect exists in the avalanche as well as streamer region ([71], [72]), no systematic analysis exists that would allow extracting the very fundamental primary information on avalanche fluctuations from the existing data.

### 6.2 Microscopic and phenomenological models

A modern approach to the avalanche problem in gaseous detectors would consist in tracking electrons in 3D in Monte Carlo fashion, considering just fundamental processes.[5] Interactions have to be, therefore, evaluated from the energy gained throughout the drift along the local electric field, together with the interaction cross-sections as a function of the former. Simulations have to be performed in connection with a field solver, provided the field needs to be simultaneously evaluated in order to properly account for the very strong avalanche space-charge that is characteristic of RPC operation. In this approach, avalanche statistics emerge naturally, but the fundamental electron cross-sections are still needed. The much increased computation time will certainly require exploring algorithms for parallel processing or introducing some type of effective/super-particle technique.

Due to its much higher simplicity, computationally-cheap phenomenological descriptions have to be considered (for examples see [74-76] and Fig. 10 for an eye-catching demonstration). As an added value, they tend to rely on general parameterizations of the system response as obtained from experimental data, incorporated into some simplified physical picture. They are focused and optimized for a particular set of observables that are critical for a certain application and they indeed do prove very useful for experiment optimization and system design (through the sometimes called ‘software digitizer’ or ‘hit producer’). Feedback towards critical technological improvements is, however, very little in this case. Either scenario (6.1, 6.2), implemented with the necessary degree of detail, might be already sufficient to describe the, as yet, totally unexplained response to photons [77], neutrons [78] and heavy ions [79, 80].

[5] The recently developed Garfield++ code [69] (maintained by CERN and developed inside the RD51 collaboration) provides a very basic and inspiring framework to do just that. As of today it is, however, not used for avalanche simulations at realistically large gains, to the best of the authors' knowledge.

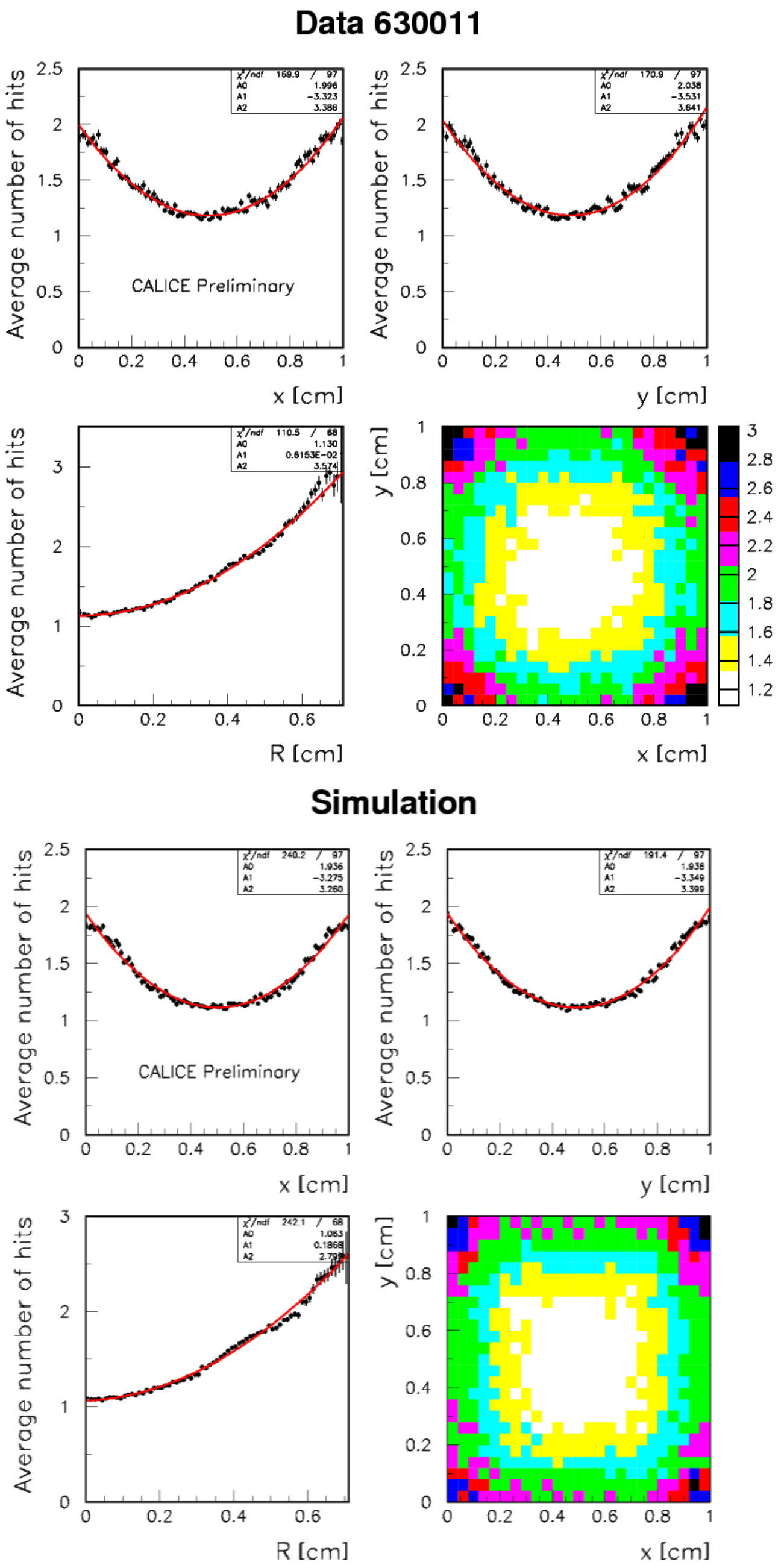

**Figure 10.** Simulation of the hit distribution within a pad and comparison with data, as obtained for the RPC-DHCAL prototype [75]. The agreement is achieved without resorting to any weighting field technique (or equivalent) for the evaluation of the induced signals. Simulations thus rely on a phenomenological parameterization of the charge distribution within the pad that, for all practical purposes, shows an excellent agreement with experimental results. *R* stands for the radial distance with respect to the center of the pad.

## 6.3 Field solvers and electromagnetic techniques

Simulations in [43] (and references therein) achieved a great degree of accuracy in the description of the complicated avalanche dynamics, which was crucial at the time, although all the necessary wrapping-up required for properly interpreting the signals registered at the end of the readout chain was treated in a rather simplified way. Fortunately, it is not difficult to improve the situation, for cases where such an additional sophistication is needed: typical RPC dimensions suggest a conceptual separation between avalanche formation and signal readout, so an 'after-burner' based on an 'induction + transmission + readout model' might suit many practical cases, by treating diverse avalanche models on equal foot [59, 65]. This framework, conceptually based on the pioneering approach sketched in W. Riegler's works [81, 82], relies on Multi-Conductor Transmission Line theory (MTL) together with a 2D electrostatic solver, and can potentially extend the 1.5D hydrodynamic framework introduced in [43] to a much higher level of realism. This is due to its ability to include practical problems connected to the system response like the characteristic impedance matrix and charge sharing phenomena (required for the description of a multi-strip/pad situation), transmission losses, connections, amplifiers and noise, without making critical approximations. In its simplest version, [65], it merely relies on 2D electrostatic parameters (that are generally much faster and easier to obtain than in the 3D case, especially for multi-strip counters). Fig. 11 shows, for illustration, a 2D induction calculation from [83] while Fig. 12 shows a transmission calculation from [59]. The latter aims at illustrating the powerful compensation scheme implemented in [84] for the case of external excitation in multi-strip RPCs. Both calculations rely on 2D finite element methods (FEM).

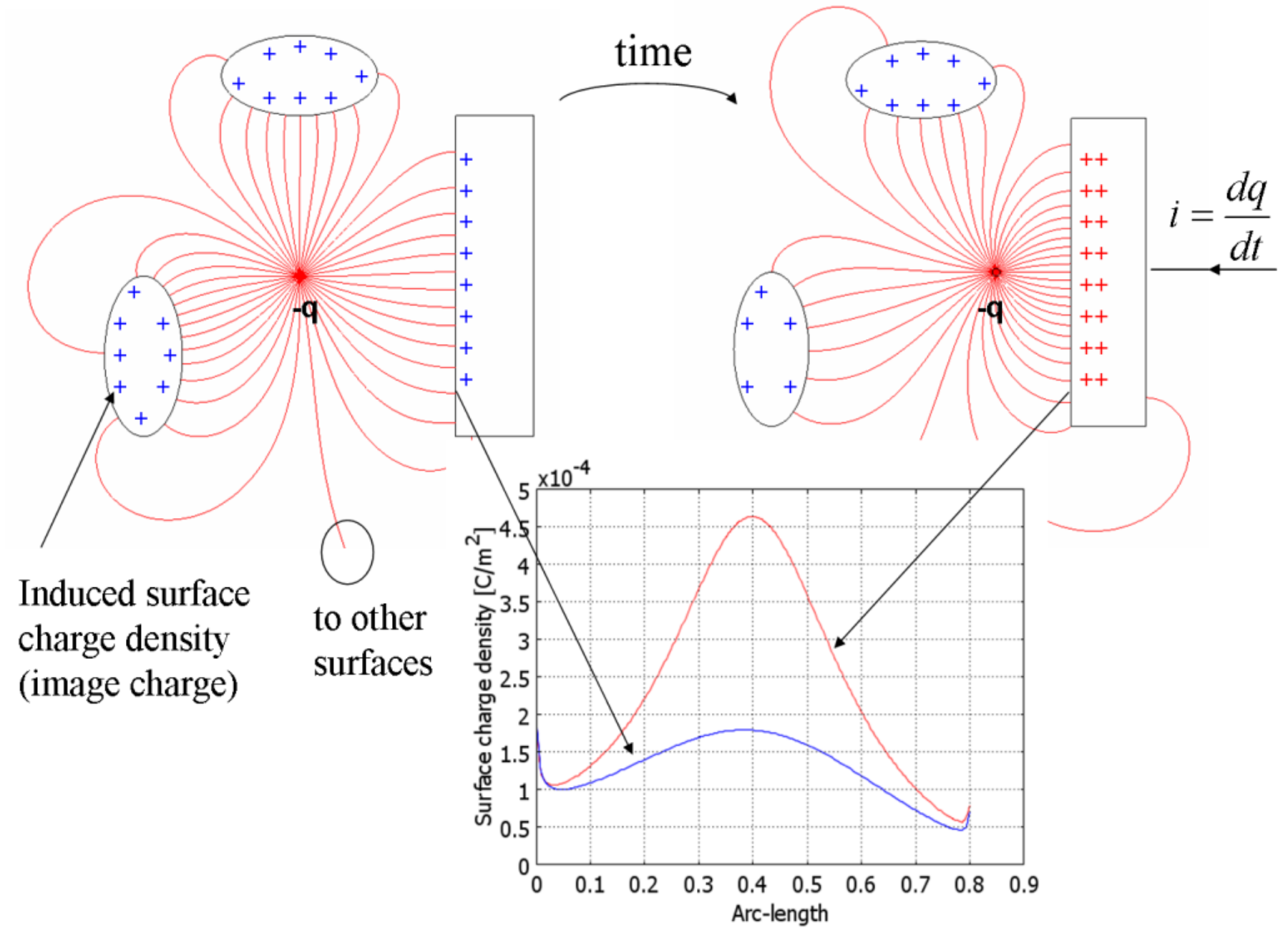

**Figure 11.** Generic induction calculation, showing the (weighting) field lines and the induced currents, from [83].

### 6.4 Towards a general 3D framework?

At last, when aiming at a detailed description of the detector response near the counter edges, spacers, but also connections as well as a completely general dynamic response of the RPC itself, it seems unavoidable to use 3D simulations in the long term. An ultimate description of the crucial multi-gap field-equilibration process needs indeed just that, an electro-dynamic (possibly quasi-static) simulation where no assumption is made on the conductivity of the gas, HV-coating, resistive plates and the spacers themselves, that would be just included in a generic RPC response function.

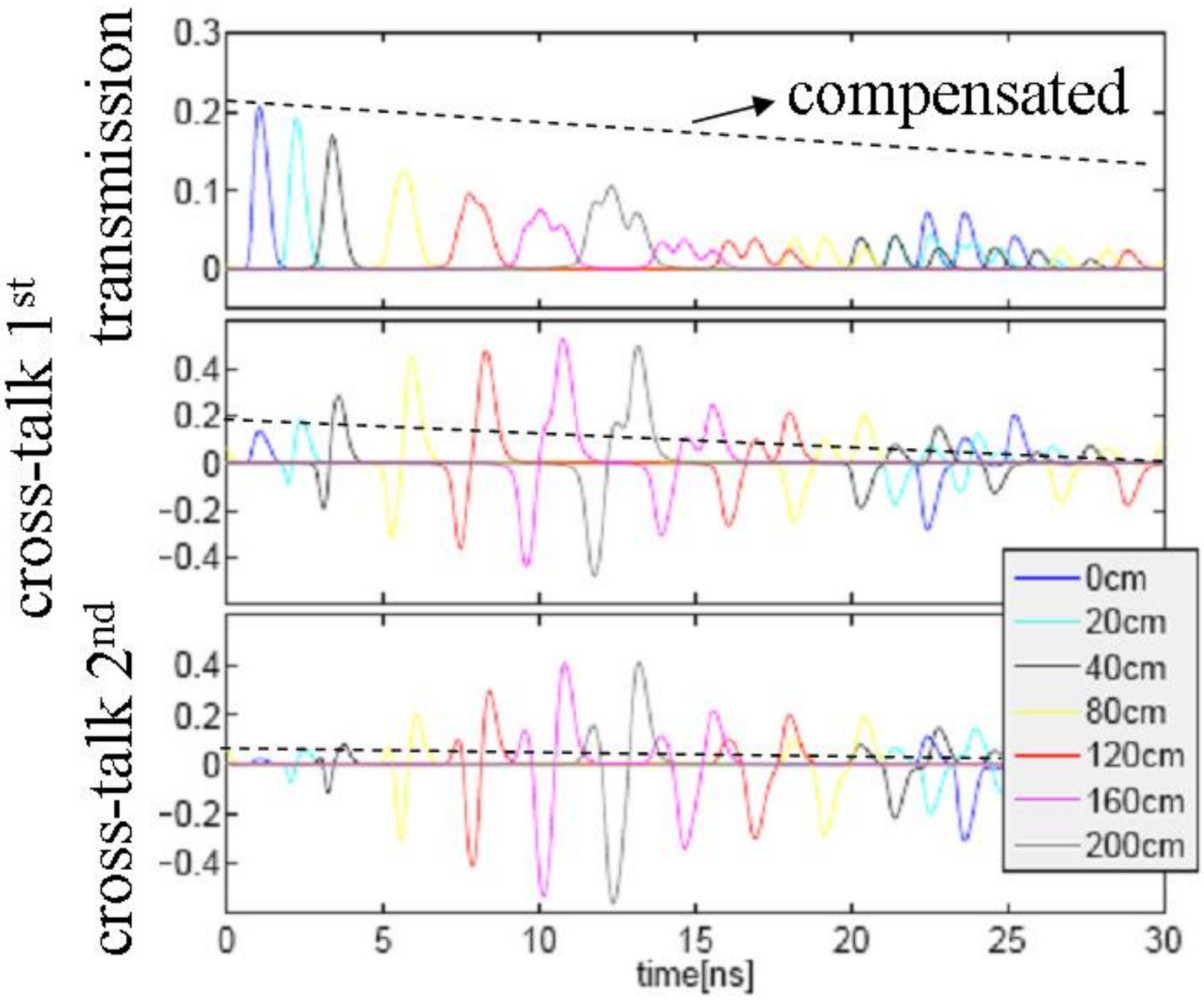

**Figure 12.** Fraction of transmitted signal and cross-talk to 1st and 2nd neighbor in a typical 2m-long RPC for different positions at which an avalanche signal has been induced, as presented at the workshop. The dashed lines represent the signal levels expected through implementation of the compensation technique introduced in [84].

## 7. Future of classic and multi-gap timing RPCs

The immediate future of classic RPCs in fundamental research is connected to the sLHC upgrade. As previously described, the ATLAS upgrade resorts to a genuine philosophy and one that reflects the large flexibility of RPCs, thanks in part to the multi-gap approach: a new challenge is addressed with a soft modification of the existing technology, thus largely minimizing possible pitfalls. The bi-gap extension, together with more sensitive electronics provides just that, an RPC capable of reaching high efficiency at 1ns resolution (or below) up to 10 kHz/cm$^2$ with a technological design very similar to that of ATLAS. For the CMS upgrade, where requirements are slightly different, there is no decision yet regarding the technological choice.

Amongst the future experiments relying on classic RPC designs, the Digital Hadron Calorimeter (DHCAL) for the International Linear Collider (ILC) features prominently: its present prototype enjoys alone almost 500.000 channels. This is allegedly causing a re-birth of digital calorimetry [75, 85]. Fig. 13 depicts a shower created by a 16 GeV/c pion as measured by the DHCAL prototype. The level of detail achieved in the description of the secondary particles is an impressive achievement and one of the strengths of the digital concept. The DHCAL design has replaced Bakelite by float glass. It will be interesting to see whether a semi-digital (multi-threshold) readout mode can really offer better performances in the challenging saturated-avalanche operation mode, and whether float glass is finally used or Chinese glass is attempted instead [86].

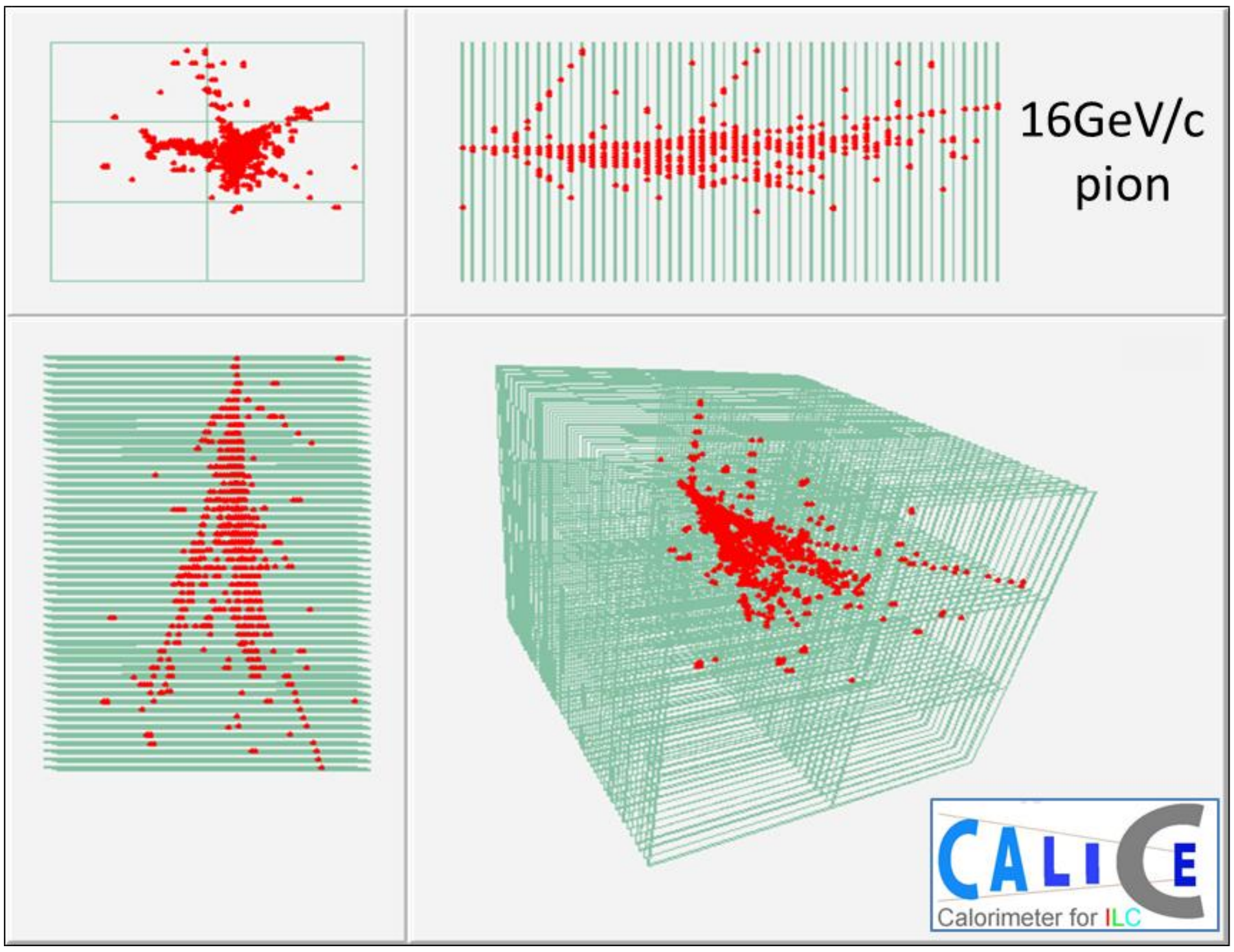

**Figure 13.** Several views of an experimentally reconstructed 16 GeV/c pion after showering on the prototype of the digital calorimeter DHCAL developed for the ILC [75].

The foreseen 28.800-module magnetized-iron neutrino telescope, ICAL, at the Indian-based Neutrino Observatory (INO) is shown in Fig. 14. Based on classic RPCs also with float glass, it will become the largest RPC-system ever [87]. Year 2012 is very important from the budgetary point of view, so there will be important news before the next RPC workshop is held.

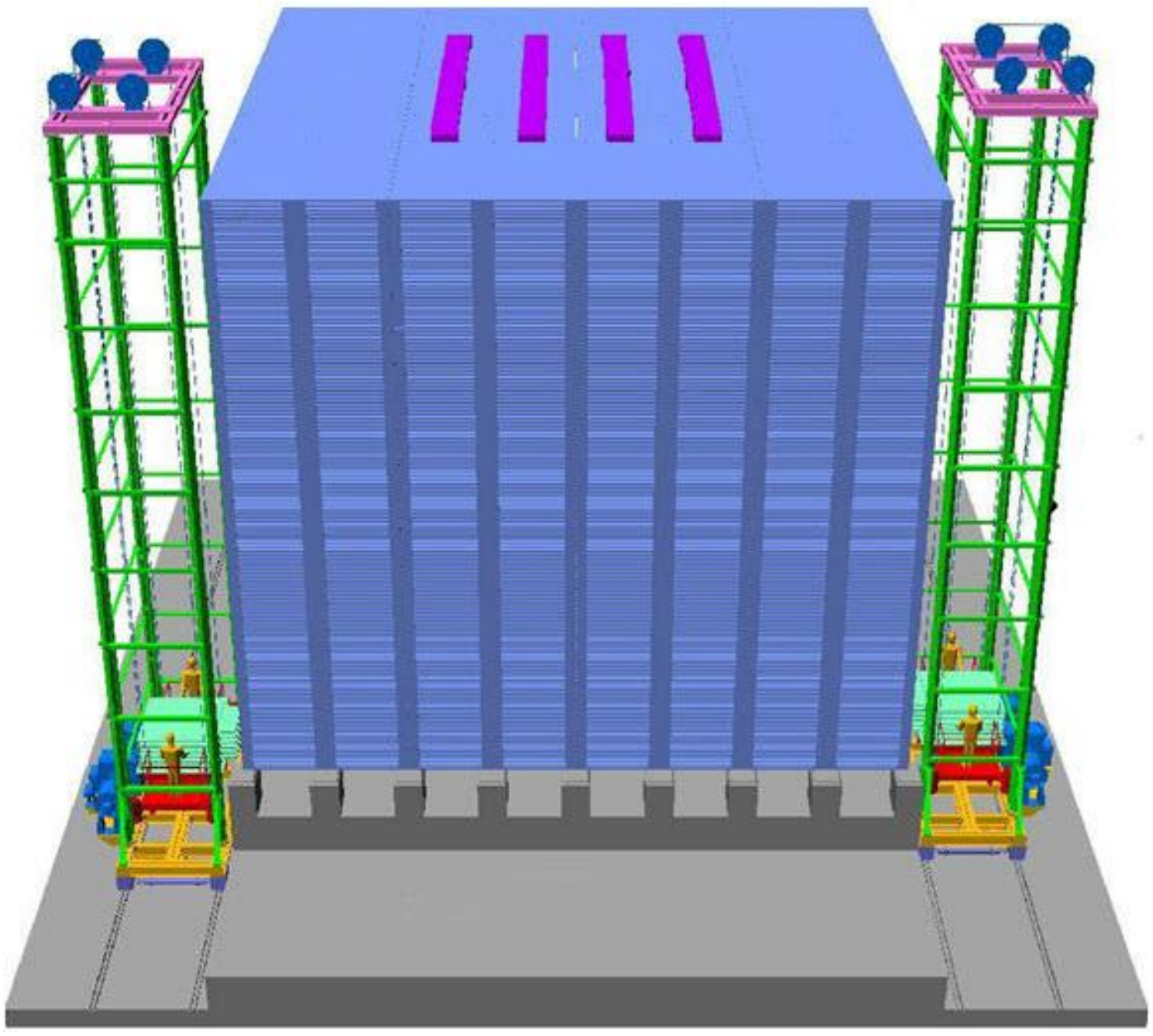

**Figure 14.** The future multi-ton India-based Neutrino Observatory. In this largely artistic view, the (hardly visible) men (in yellow) at left and right are participating on the system assembly.

Concerning the usage of multi-gap timing RPCs in fundamental physics, future developments are largely circumscribed to the Facility for Antiproton and Ion Research (FAIR), Darmstadt, Germany. There, two experiments, the Compressed Baryonic Matter experiment (CBM) as well as $R^3B$ (Reactions with Relativistic Radioactive Beams) are planning to make use of MRPC walls. There are several ideas that will be worth following: the Chinese glass - based modules [41, 45] and the evolution of the 'warm glass technique' [88, 89] aimed at a ~x10 rate capability increase (Fig. 15), as well as the differential strip readout [90, 91], whose technological relevance has still to be clarified. It will be also an opportunity to see if it is really possible to obtain good system performances at low levels of inter-strip cross-talk, as some dedicated measurements suggest [84]. This is of paramount importance for the CBM-type heavy ion experiments, where particle multiplicity can be dramatically high. Spurious cross-talk or electromagnetic pick-up can easily jeopardize the challenging 80ps system-resolution that is envisaged [91].

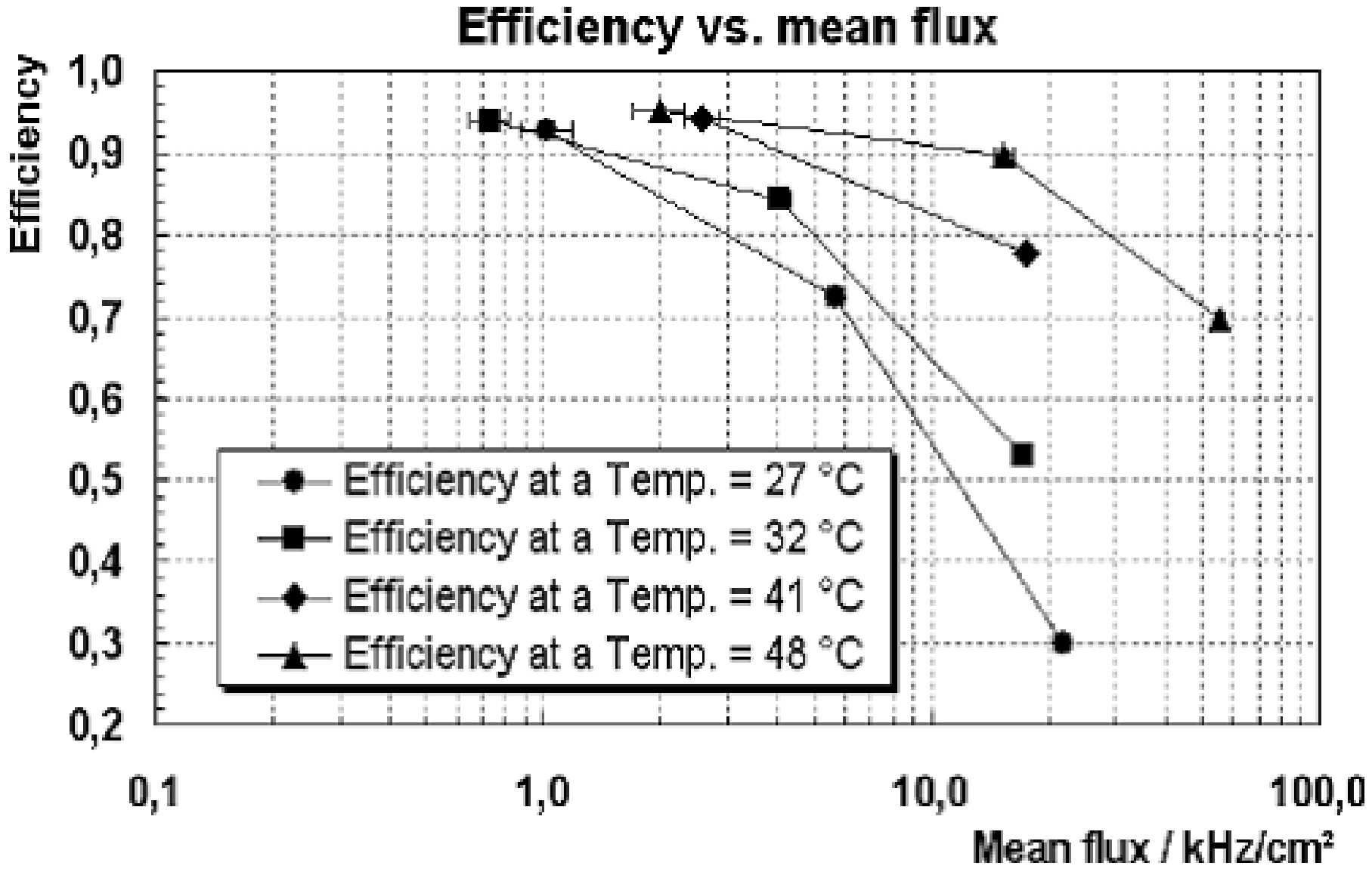

**Figure 15.** Efficiency as a function of the average particle flux impinging over the counter for CBM prototypes based on standard float glass but resorting to the 'warm glass technique' (taken from [89]). The technique was first demonstrated for timing applications in [88].

On the other hand, the general-purpose $R^3B$ experiment aims at detecting neutrons and heavy ions with two different setups, both with sub-100ps time resolution. Although the first system is almost surely discontinued in virtue of a scintillator calorimeter (see [92], however), it will be very interesting to see the evolution of the latter. Dedicated studies of the detector response under heavy ions are scarce [79, 80, 93] and they will undoubtedly shed light on the detector behavior, specifically on the space charge phenomena and the ionization quenching factors for highly ionizing particles, that are totally unknown for RPC-mixtures as of today. For illustration, the RPC time response for the heaviest nuclei studied to date (Xe) is shown in Fig. 16 at different fluxes [80].

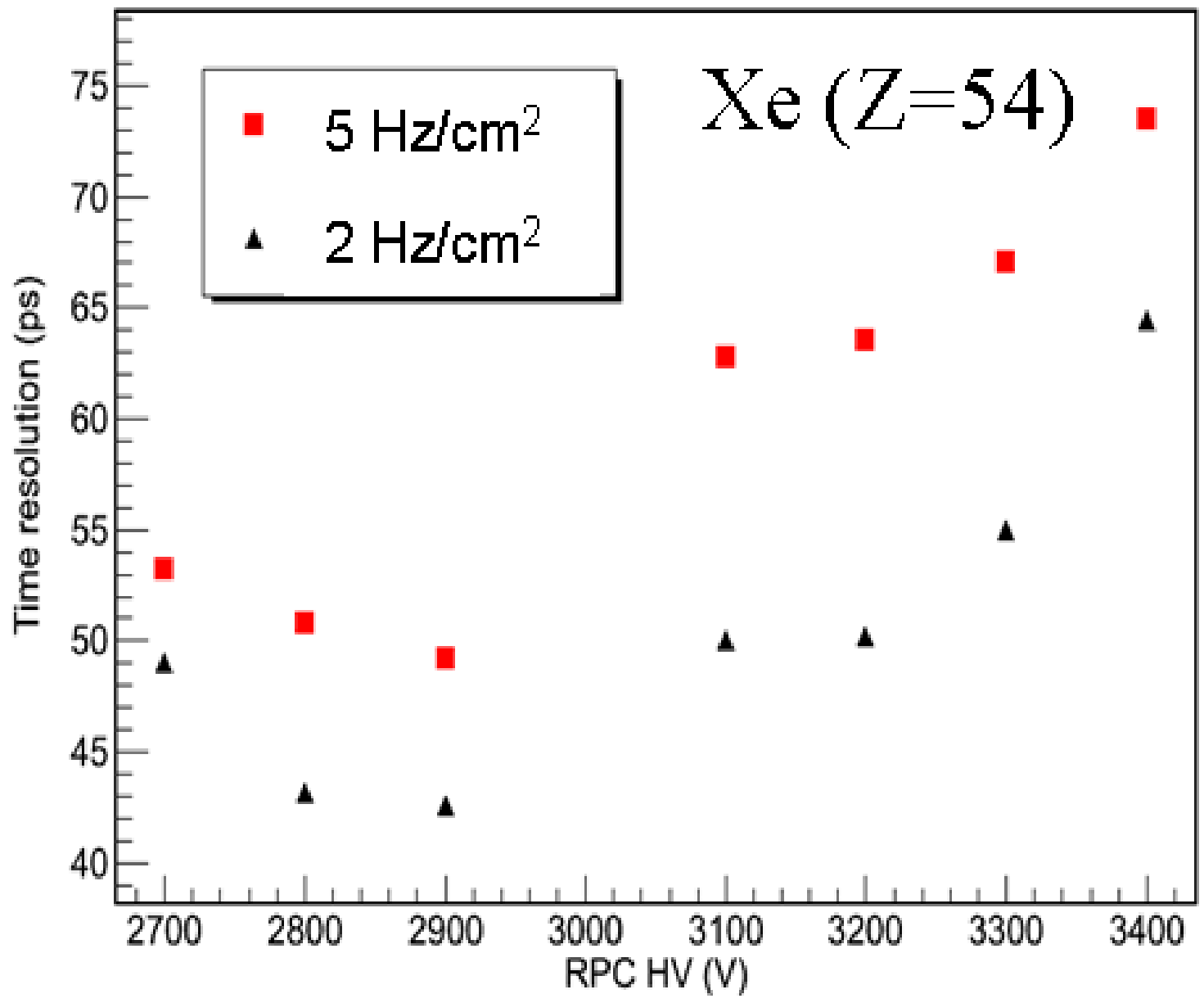

**Figure 16.** RPC time response for Xenon ions at different ion flux, as part of the $R^3B$ R&D (adapted from [80]).

Besides FAIR, another facility to look at is Spring, Japan, and in particular the new Laser Electron Photon experiment. Its aim is studying photon-induced hadro-production, for which a high-end 50ps resolution over $5m^2$ is required, in order to provide sufficient $\pi$/k separation. In view of the low granularity, a multi-strip configuration in two layers seems to comfortably satisfy the requirements, but at the moment all options for the electrode layout seem to be open [94].

## 8. New applications

The scope of future applications remains rather constrained to large-area tomography with high position resolution. Applications are largely focused on Positron Emission Tomography (PET), following the pioneering work in [54], and so at present there are several groups involved on different concepts for its optimization [95, 96]. The RPC-PET makes critical use of the time information in order to improve the position reconstruction along the 2-photon annihilation-line, as well as of the position resolution (in order to reconstruct the position perpendicular to the annihilation-line) and so it represents an excellent platform for further developing the technology (Fig. 17-left). In this context, understanding the worse resolution obtained for annihilation photons (90-100ps) as compared to minimum ionizing particles will be undoubtedly an important break-through in order to seek further optimization [77, 88].

Muon tomography for cargo screening and general border monitoring of dense materials, on the other hand, is well suited to the RPC technology and can be already accomplished to the accuracy needed with state of the art classic RPCs [97], without resorting to extreme timing capabilities (Fig. 17-right).

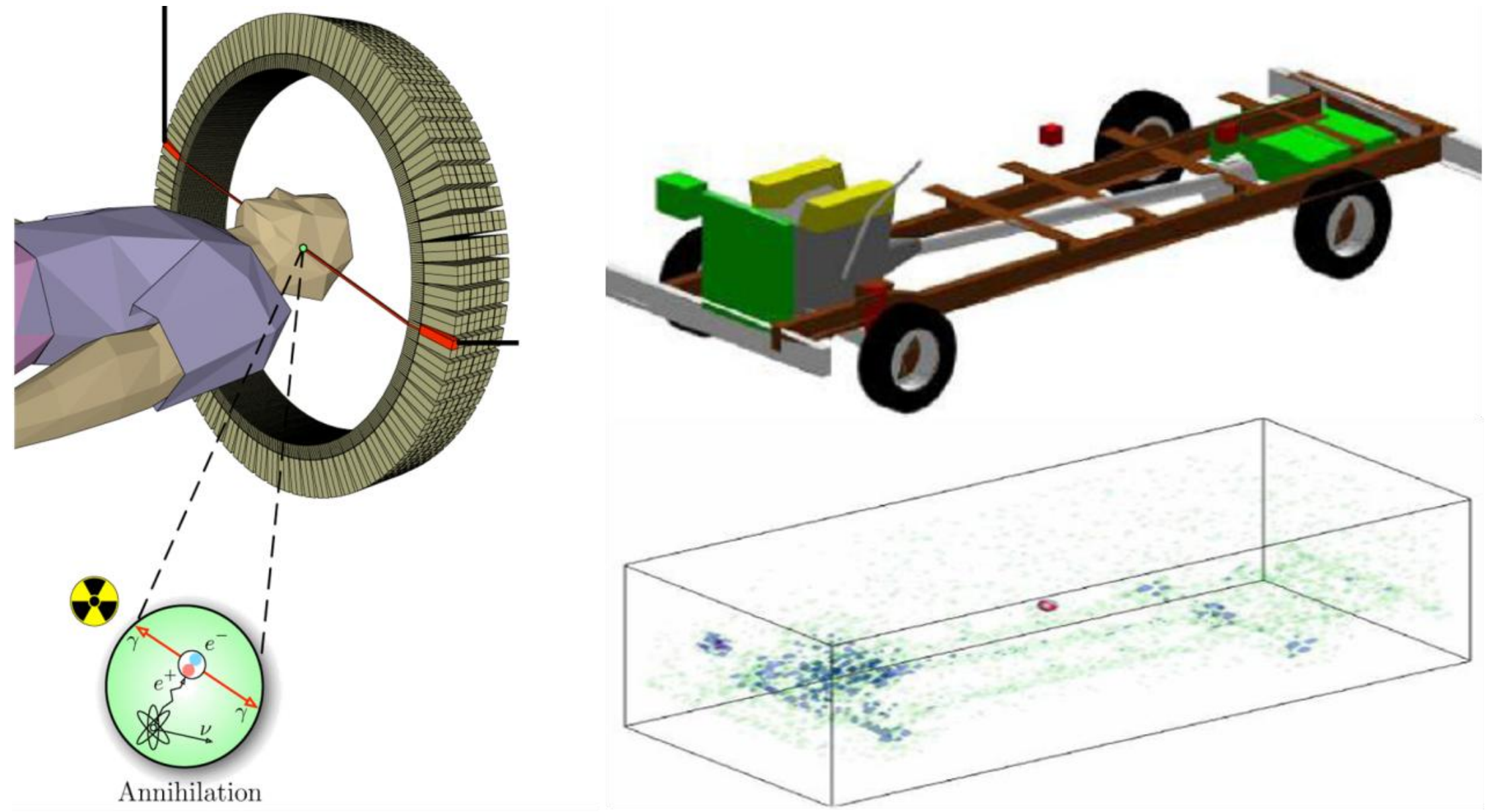

**Figure 17.** Left: General sketch of the PET concept. The position resolution along the 2-photon annihilation-line is around 3cm for a 100ps time-difference resolution (characteristic of multi-gap timing RPCs), improving by 3mm every additional 10ps.  Right: simulation results obtained by cosmic muon tomography using classic RPCs. The 10x10x10cm$^3$ (red) tungsten block can be found in 1 minute (from [97]).

A very interesting tracking-timing technology was introduced at the workshop [95], by performing timing and tracking with a single detector (better than 100μm×100μm space and 100ps time resolutions were simultaneously achieved). This new idea matches quite naturally the recent proposal for doing 4D-tracking including time information [99]. Although promising, its final relevance for tracking and timing will depend on considerations regarding material budget, system occupancy and especially finding a window of opportunity for the 4D-tracking idea in terms of an adequate physical environment, geometry, and detector performance. In other words, 4D-tracking is expected to bring a significant performance improvement as compared to conventional 3D-tracking in environments where: i) event pile-up (time-wise) is high and/or ii) the typical time-of-flight spread over tracking stations is significantly larger than the overall time-of-flight accuracy of the system [99].

Despite the involvement of several groups in these activities, the impact of RPCs in everyday life is still limited and new ideas are necessary.

## 9. New trends on Resistive Gaseous Detectors

Resistive Gaseous Detectors quench spark formation down to ‘localized discharges’ [4], generally believed to be of the streamer-type in most of the so far existing practical applications ([100], for instance). The locality of the process is achieved by limiting the current flow over the electrode surface. There is nothing special about parallel plate geometries that makes them especially suited to this technique except perhaps that they are often forced to work under very high fields in order to achieve the fastest possible timing response. The growth of solid state alternatives that usually offer a higher reliability as compared to gaseous detectors might explain the changing trend but, either way, some of the high-end micro-patterned gaseous detectors are currently developed following insulating techniques known to the RPC field years ago. The insulated design offers a much improved detector robustness at the price of a reduced rate capability (due to the charge built-up on the insulator), however a practical compromise can be often achieved.

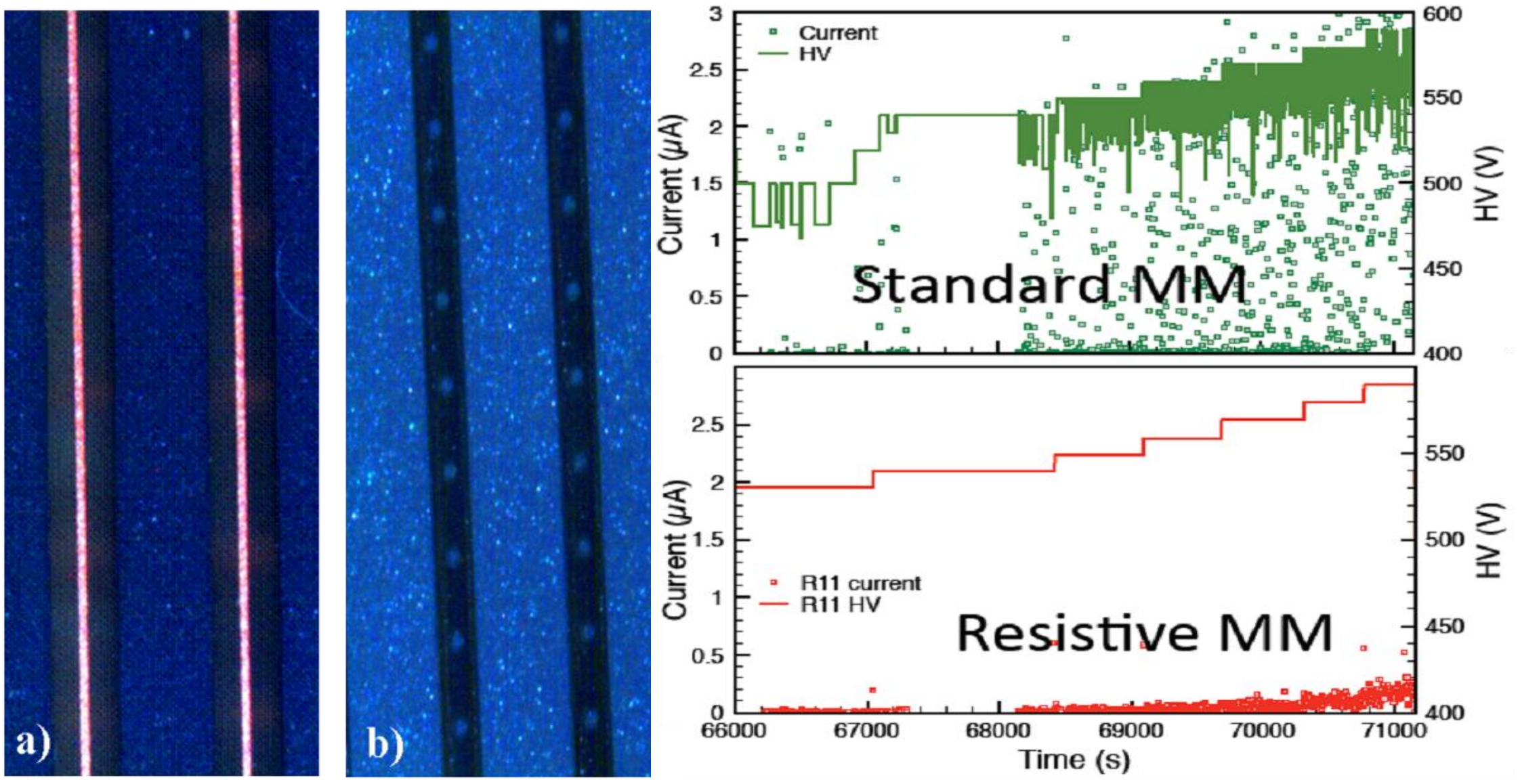

**Figure 12:** Left: a) A resistive Micro-Strip Gas Detector (anode is conductive) and b) a fully resistive Micro-Dot Gas Detector (adapted from [101]). Right: comparison between standard and resistive Micromegas (MM), from [102].

Ordinary micro-pattern gaseous detectors (MSGC, Micromegas, GEM and others; -see [101] and references therein) can be seen as proportional counters offering unprecedented 2D position resolution (20-40 μm) and are thus very attractive for many applications: tracking of charged particles, photon-imaging, etc. It has been demonstrated by the RD51 community that the resistive electrode approach can be successfully applied to micro-pattern gaseous detectors and make them robust and spark-protected. Latest designs of resistive micro-pattern detectors have their electrodes segmented with resistive strips (Fig. 12-left). This approach turned out to be very fruitful and has allowed developing different designs of large-area spark-protected micro-pattern detectors oriented to various applications. Perhaps most remarkably, Micromegas with resistive anode strips are envisaged for the 1000 $m^2$ forward wheel upgrade of ATLAS [102]. The dramatic improvements in terms of spark suppression and general stability are well illustrated by Fig. 12-right. Since developments aimed at high position resolution are already existing for the, so called, micro-strip RPCs [103] it seems unavoidable to have in the near future resistive micro-pattern gaseous detectors and genuine RPCs that are functionally equivalent for certain applications. It will be very interesting to see how the technological scenario evolves from that point on.

## 10. Conclusions

We have presented the aims and ambitions of the RPC community towards the next international workshop to be held in Beijing in 2014, and we have done it from the general point of view of resistive gaseous detectors of arbitrary geometry. Seemingly, many of the goals can be achieved in the 2-year time interval spanning between workshops. Amongst them, the consolidation of most of the already running and upcoming systems, new materials and high-rate techniques together with improved avalanche simulations and next-generation front-end electronics. Many others, however, like both streamer and generic 3D electro-dynamic simulations, the search for new gases, and the development of general-purpose front-end electronics, might develop at a slower pace. Above all, the field clearly lives some exciting times at the moment, and this will certainly continue materializing in new applications and ideas. We look forward to 2014.

## Acknowledgments

DGD is partially supported by the National Science Foundation of China, under grant 111050110573. DGD wants to acknowledge support and discussions with colleagues, among them, M. Ciobanu, W. Riegler, P. Fonte, R. Santonico, R. Cardarelli, V. Peskov, M. Abbrescia, I. Iomataris, I. Irastorza and Y. Wang.